\documentclass[aps,prd,amsmath
,nofootinbib         
]{revtex4}
\usepackage{amsmath}
\usepackage{amsfonts}
\usepackage{amssymb}

\allowdisplaybreaks[4]     

\begin{document}

\title{Stationary cylindrical anisotropic fluid and new purely magnetic GR solutions}

\author{Marie-No\"{e}lle C\'{e}l\'{e}rier}
\affiliation{Observatoire de Paris-Meudon, 5, Place Jules Janssen, F-92195 Meudon Cedex, France}
\email{marie-noelle.celerier@obspm.fr}

\author{Nilton O. Santos}
\affiliation{Sorbonne Universit\'e, UPMC Universit\'e Paris 06, LERMA, UMRS8112 CNRS,\\
Observatoire de Paris-Meudon, 5, Place Jules Janssen, F-92195 Meudon Cedex, France}
\email{Nilton.Santos@obspm.fr}

\date{April 2 2020}

\begin{abstract}

The properties of interior spacetimes sourced by stationary cylindrical anisotropic fluids are here analytically studied for both nonrigid and rigid rotation. As regards nonrigid rotation, this is, to our knowledge, the first work dedicated to such a study. We give here a complete equation set describing these spacetime properties. In particular, we focus our attention on both nonrigid and rigid rotation gravito-electromagnetic features and are thus led to display strong hints in favor of conjecturing purely  electric Weyl tensor existence in this framework. We have also been able to characterize new purely magnetic physically consistent spacetimes and have found new rigidly rotating exact solution classes to the five Einstein's field equations pertaining to the issue and the two purely magnetic constraints we have derived for this purpose. This should be considered as a prominent result, since extremely few purely magnetic exact solutions are available in the literature. 
\end{abstract}

\maketitle

\section{Introduction} \label{intro}

A long-standing technique to assist in spacetime metric studies involving Einstein's general relativity field equations is to impose symmetry constraints, i.e., Killing vectors. Now, one Killing vector implies a rather tricky problem, while with two, even though still involved, the problem simplifies such as to become rather tractable, at least in a number of simplifying cases \cite{G09,S09}. Cylindrical symmetry, implying two Killing vectors, has therefore attracted much attention since the pioneering work by Levi-Civita identifying in 1919 vacuum static cylindrical spacetimes \cite{LC19} and their extension to stationary ones obtained independently by Lanczos in 1924 \cite{L24} and by Lewis in 1932 \cite{L32}. The Lewis solution describes a vacuum exterior sourced by a matter cylinder rotating around its symmetry axis. The vacuum solution outside a cylindrical source in translation along its symmetry axis is mathematically akin to the Lewis solution with exchanged $z$ and $\phi$ coordinates. It has been shown that they are, however, physically different \cite{C19}. Nonvacuum cylindrically symmetric spacetime investigations date back to 1937 when van Stockum gave the metric solution for a rigidly rotating infinitely long dust cylinder \cite{vS37}. Since then, cylindrically symmetric spacetimes have been extensively investigated for a number of different purposes \cite{G09,S09}. For a recently published review on cylindrical systems in general relativity, see \cite{B20}.

In \cite{D06}, nonvacuum stationary spacetimes sourced by a cylindrical anisotropic fluid have been considered, while only rigid rotation has been studied. In the present work, we extend this study to the nonrigid rotation case, and complete and improve the rigid case analysis proposed in \cite{D06}.  We first aim here at thoroughly analyzing the stationary nonrigidly rotating anisotropic fluid cylinder mathematical and physical features, with a focus on its Weyl tensor gravito-electromagnetic properties. Then we exemplify our results by applying the obtained equation set to the rigid rotation case. The more prominent result we obtain in the gravito-electromagnetism framework is the display of purely magnetic Weyl tensor spacetimes exhibiting physically consistent properties.

Actually, any nonconformally flat spacetime's Weyl tensor can be pointwise decomposed into an electric, $E_{\alpha \beta}$, and a magnetic, $H_{\alpha \beta}$, part with respect to a given unit timelike congruence, $u^\alpha$, and this decomposition determines entirely the Weyl tensor. $E_{\alpha \beta}$ and $H_{\alpha \beta}$ are traceless, symmetric and spacelike tensors. Such a decomposition was first introduced in \cite{M53} and developed in \cite{B62}, where it has been applied to the vacuum Riemann tensor in an attempt to find out a possible analogy between gravitational and classical electromagnetic quantities. However, such an analogy is not that straightforward, and can even be physically misleading if not handled with care \cite{C14}.

A nonconformally flat spacetime for which the Weyl tensor magnetic (electric) part identically vanishes with respect to some $u^\alpha$ is called purely electric (magnetic) and its Weyl tensor is said to be purely electric (magnetic) with respect to $u^\alpha$. Since  $E_{\alpha \beta}$ and $H_{\alpha \beta}$ (with respect to $u^\alpha$) are diagonalizable, the purely electric (magnetic) spacetime Petrov type is necessarily $I$ or $D$, and, at each point, $u^\alpha$ is a Weyl principal vector and is essentially unique (up to sign) for the Petrov type $I$ spacetimes \cite{L02,B04} which will be considered here. Now, the purely electric (magnetic) types can be characterized with no reference to $u^\alpha$ by using the complex quadratic, cubic and zero-dimensional Weyl tensor invariants. 

Actually, while many purely electric spacetimes are known \cite{B89,L95,BMS96,vE97,M99,F01,V04,V05,W06a} and, in particular, all static ones, this is not the case for purely magnetic spacetimes even though there can be found in the literature a rather great involvement in trying to find out mathematically and/or physically consistent such solutions \cite{L02,B04,A94,B95,M98,F99,LA99,LM99,L03,W06b,L07}. Currently, no purely magnetic vacuum solutions, with or without $\Lambda$ cosmological constant, are known. This has led to the conjecture that purely magnetic vacua do not exist in an open four-dimensional (4D) region \cite{B04,M94}. Proofs for this conjecture have been displayed, but merely for spacetimes exhibiting particular Petrov types or specific physical properties \cite{M94,T65,B75,H95,V03a,V03b,F03,F04a,F04b,Z05}. As regards nonvacuum solutions, almost every known purely magnetic spacetimes are Petrov type $D$ \cite{L02,F99,LA99,L03,S68,vE96,B96,Ma97,Ma99} or $I(M^\infty)$ in the extended McIntosh-Arianrhod classification \cite{M90,A92}, and the last one only for perfect fluids \cite{B04}. The only type $I(M^+)$ purely magnetic spacetime available in the literature has been mathematically constructed and its physical meaning remains unspecified \cite{L07}. The specific improvement here proposed consists in obtaining $I(M^\infty)$ purely magnetic spacetimes sourced by actual physically consistent fluids of which we give a detailed analysis. We have thus been led to exhibit rigidly rotating exact solutions to the whole set of five Einstein's field equations pertaining to the issue and the two purely magnetic constraint equations we have derived for the purpose.

Even though Hawking and Ellis \cite{H73} have excluded purely magnetic spacetimes from their singularity theorem conditions, arguing that such spacetimes should be unphysical, their argument applies only for modeling the whole universe. There is no obvious physical reason why purely magnetic solutions such as the ones displayed here should not be used as models for particular astrophysical objects or for some spacetime regions.

The paper is organized as follows: in Sec. \ref{ci}, we set up the stationary cylindrically symmetric line element which will be used for both nonrigidly rotating and rigidly rotating fluid classes. In Sec. \ref{nrrf}, we display the field equations, the hydrodynamical scalars, vectors and tensors, the regularity and junction conditions, and a gravito-electromagnetic analysis conducted for the nonrigid rotation case. We are thus led to propose, as a strongly based conjecture, that purely electric and purely magnetic Petrov type $I(M^\infty)$ such nonstatic spacetimes exist and we give the simplified equation set they verify in each case, purely electric or purely magnetic. Section \ref{rr} is devoted to an analogous rigid case analysis. We show that purely electric cylindrically symmetric spacetime staticity can merely be {\it conjectured}, while {\it not proved}, for rigid rotation. Purely magnetic rigidly rotating spacetimes sourced by a fluid with vanishing radial and azimutal stresses are given in Sec. \ref{example} as exact solutions for both the field equation set and the purely magnetic constraints derived in Sec. \ref{rr}. Our conclusions are displayed in Sec. \ref{concl}.

\section{Cylindrical spacetime inside the source} \label{ci}

We consider a stationary cylindrically symmetric anisotropic nondissipative fluid bounded by a cylindrical surface $\Sigma$ and whose stress-energy tensor we write as
\begin{eqnarray}
T_{\alpha \beta} = (\rho + P_r) V_\alpha V_\beta + P_r g_{\alpha \beta} + (P_\phi - P_r) K_\alpha K_\beta + (P_z - P_r)S_\alpha S_\beta, \label{setens}
\end{eqnarray}
where $\rho$ is the fluid energy density; $P_r$, $P_z$, and $P_\phi$ are the principal stresses; and $V_\alpha$, $K_\alpha$, and $S_\alpha$ are 4-vectors satisfying
\begin{eqnarray}
V^\alpha V_\alpha = -1, \quad K^\alpha K_\alpha = S^\alpha S_\alpha = 1, \quad V^\alpha K_\alpha = V^\alpha S_\alpha = K^\alpha S_\alpha =0. \label{fourvec}
\end{eqnarray}
We assume, for the inside $\Sigma$ spacetime, the spacelike $\partial_z$ Killing vector to be hypersurface orthogonal, such as to ease its subsequent matching to the exterior Lewis metric Weyl class. Hence, the stationary cylindrically symmetric line element reads
\begin{equation}
\textrm{d}s^2=-f \textrm{d}t^2 + 2 k \textrm{d}t \textrm{d}\phi +\textrm{e}^\mu (\textrm{d}r^2 +\textrm{d}z^2) + l \textrm{d}\phi^2, \label{metric}
\end{equation}
where $f$, $k$, $\mu$, and $l$ are real functions of the radial coordinate $r$ only. Owing to cylindrical symmetry, the coordinates are bound to conform to the following ranges:
\begin{equation}
-\infty \leq t \leq +\infty, \quad 0 \leq r, \quad -\infty \leq z \leq +\infty, \quad 0 \leq \phi \leq 2 \pi, \label{ranges}
\end{equation}
where the two limits of the $\phi$ coordinate are topologically identified. We number the coordinates $x^0=t$, $x^1=r$, $x^2=z$, and $x^3=\phi$.

\section{Nonrigidly rotating fluid} \label{nrrf}

We consider first the nonrigid rotation case. The fluid 4-velocity, satisfying conditions (\ref{fourvec}), can thus be chosen as
\begin{equation}
V^\alpha = v \delta^\alpha_0 + \Omega \delta^\alpha_3, \label{nr4velocity}
\end{equation}
where $v$ and $\Omega$ are functions of $r$ only. Since $V^\alpha$ has to satisfy the timelike condition provided in (\ref{fourvec}), we have
\begin{equation}
fv^2 - 2kv\Omega - l \Omega^2 -1 = 0. \label{timelike}
\end{equation}
The two spacelike 4-vectors used to define the stress-energy tensor, while verifying conditions  (\ref{fourvec}), can be chosen as
\begin{equation}
K^\alpha = -\frac{1}{D}\left[(kv+l\Omega)\delta^\alpha_0 + (fv - k\Omega)\delta^\alpha_3\right], \label{kalpha}
\end{equation}
\begin{equation}
S^\alpha = \textrm{e}^{-\mu/2}\delta^\alpha_2, \label{salpha}
\end{equation}
with
\begin{equation}
D^2 = fl + k^2. \label{D2}
\end{equation}

\subsection{Field equations} \label{fe}

With the above choice for the three 4-vectors defining the stress-energy tensor, and using (\ref{timelike}) into (\ref{setens}), we obtain the five nonzero components of this stress-energy tensor, and we can write the following five Einstein's field equations for the inside $\Sigma$ spacetime:
\begin{eqnarray}
G_{00} = \frac{\textrm{e}^{-\mu}}{2} \left[-f\mu'' - 2f\frac{D''}{D} + f'' - f'\frac{D'}{D} + \frac{3f(f'l' + k'^2)}{2D^2}\right] = \kappa \left[\rho f + (\rho + P_\phi)D^2\Omega^2\right], \label{G00}
\end{eqnarray}
\begin{eqnarray}
G_{03} =  \frac{\textrm{e}^{-\mu}}{2} \left[k\mu'' + 2 k \frac{D''}{D} -k'' + k'\frac{D'}{D} - \frac{3k(f'l' + k'^2)}{2D^2}\right] = - \kappa \left[\rho k + (\rho + P_\phi)D^2 v \Omega\right], \label{G03}
\end{eqnarray}
\begin{eqnarray} 
G_{11} = \frac{\mu' D'}{2D} + \frac{f'l' + k'^2}{4D^2} = \kappa P_r \textrm{e}^\mu, \label{G11}
\end{eqnarray}
\begin{eqnarray}
G_{22} = \frac{D''}{D} -\frac{\mu' D'}{2D} - \frac{f'l' + k'^2}{4D^2} = \kappa P_z \textrm{e}^\mu, \label{G22}
\end{eqnarray}
\begin{eqnarray}
G_{33} =  \frac{\textrm{e}^{-\mu}}{2} \left[l\mu'' + 2l\frac{D''}{D} - l'' + l'\frac{D'}{D} - \frac{3l(f'l' + k'^2)}{2D^2}\right] = - \kappa \left[\rho l - (\rho + P_\phi)D^2 v^2\right], \label{G33}
\end{eqnarray}
where the primes stand for differentiation with respect to $r$.

We have thus six equations, i. e., (\ref{timelike}) and (\ref{G00})-(\ref{G33}) for ten unknown functions of $r$, namely, $f$, $k$, $\mu$, $l$, $v$, $\Omega$, $\rho$, $P_r$, $P_z$, and $P_\phi$. Therefore, four equations of state connecting the matter observables or {\it ad hoc} assumptions on the metric  functions would have to be imposed in order to solve the field equations. However, (\ref{G00})-(\ref{G33}) can be partially integrated as follows. From (\ref{G00}) and (\ref{G33}), we can derive
\begin{equation}
\left(\frac{lf' - fl'}{D}\right)' = 2 \kappa (\rho + P_\phi) D \textrm{e}^\mu (fv^2 + l\Omega^2). \label{partint1}
\end{equation}
From  (\ref{G00}) and (\ref{G03}), we obtain
\begin{equation}
\left(\frac{kf' - fk'}{D}\right)' = 2 \kappa (\rho + P_\phi) D \textrm{e}^\mu (k\Omega^2 - fv\Omega). \label{partint2}
\end{equation}
Equations (\ref{G03}) and (\ref{G33}) yield
\begin{equation}
\left(\frac{kl' - lk'}{D}\right)' = - 2 \kappa (\rho + P_\phi) D \textrm{e}^\mu (kv^2 + lv\Omega). \label{partint3}
\end{equation}
Using  (\ref{partint1})-(\ref{partint3}), and assuming $\Omega \neq 0$, i.e., the nonrigid rotation case, with $fv^2 \neq - l \Omega^2$, $k\Omega \neq fv$, $kv \neq -l\Omega$, and, of course, $v \neq 0$, we can write
\begin{eqnarray}
\frac{1}{fv^2 + l\Omega^2}\left(\frac{lf' - fl'}{D}\right)' = \frac{1}{k\Omega^2 - fv\Omega}\left(\frac{kf' - fk'}{D}\right)' = \frac{1}{kv^2 + lv\Omega}\left(\frac{lk' - kl'}{D}\right)'. \label{partint}
\end{eqnarray}

\subsection{Hydrodynamical scalars, vectors, and tensors} \label{tobetitled}

The timelike 4-vector $V_\alpha$ can be invariantly decomposed into three independent parts through the genuine tensor $V_{\alpha;\beta}$ as
\begin{equation}
V_{\alpha;\beta} = - \dot{V}_\alpha V_\beta + \omega_{\alpha \beta} + \sigma_{\alpha \beta}, \label{decomp}
\end{equation}
where 
\begin{equation}
 \dot{V}_\alpha = V_{\alpha;\beta} V^\beta, \label{accel}
\end{equation}
\begin{equation}
\omega_{\alpha \beta} = V_{\left[\alpha;\beta\right]} +  \dot{V}_{\left[\alpha\right.} V_{\left.\beta\right]}, \label{rotation}
\end{equation}
\begin{equation}
\sigma_{\alpha \beta} = V_{(\alpha;\beta)} +  \dot{V}_{(\alpha} V_{\beta)}. \label{shear}
\end{equation}
The three above quantities are called, respectively, the acceleration vector, the rotation or twist tensor, and the shear tensor. For the timelike 4-vector given by (\ref{nr4velocity}), the (\ref{accel})-(\ref{shear}) nonzero components are
\begin{equation}
 \dot{V}_1 = - \Psi, \label{dotV1}
\end{equation}
\begin{equation}
2 \omega_{01} = - (fv-k\Omega)' - (fv-k\Omega)\Psi, \label{omega01}
\end{equation}
\begin{equation}
2 \omega_{13} = - (kv+l\Omega)' - (kv+l\Omega)\Psi, \label{omega13}
\end{equation}
\begin{equation}
2 \sigma_{01} = - fv' + k\Omega' + (fv - k\Omega) \Psi, \label{sigma01}
\end{equation}
\begin{equation}
2 \sigma_{13} = kv' + l\Omega' - (kv + l\Omega) \Psi, \label{sigma13}
\end{equation}
with
\begin{equation}
\Psi = fvv' - k(v\Omega)' - l\Omega\Omega' = -\frac{1}{2} (v^2f' - 2v\Omega k' - \Omega^2l'), \label{psidef}
\end{equation}
where the equality in (\ref{psidef}) follows from (\ref{timelike}) differentiated with respect to $r$.
The modulus of the acceleration vector is
\begin{equation}
\dot{V}^\alpha \dot{V}_\alpha =  \textrm{e}^{-\mu} \Psi^2. \label{modaccel}
\end{equation}
The rotation scalar, $\omega$, defined by
\begin{equation}
\omega^2 = \frac{1}{2} \omega^{\alpha \beta}\omega_{\alpha \beta}, \label{omega2def}
\end{equation}
follows from
\begin{eqnarray}
\omega^2 = \frac{f}{4  \textrm{e}^{\mu}D^2(fv - k\Omega)^2} \left[(kf' - fk')v^2 + (lf' - fl') v \Omega + (kl' - lk') \Omega^2 + D^2(v'\Omega - v\Omega')\right]^2. \label{omega2}
\end{eqnarray}
The shear scalar, $\sigma$, defined by
\begin{equation}
\sigma^2 = \frac{1}{2} \sigma^{\alpha \beta}\sigma_{\alpha \beta}, \label{sigma2def}
\end{equation}
follows from 
\begin{equation}
\sigma^2 = \frac{\textrm{e}^{-\mu}D^2}{4}(v'\Omega - v\Omega')^2 . \label{sigma2} 
\end{equation}
Equation  (\ref{sigma2}) shows that choosing a corotating frame for the stationary fluid source, implying $\Omega = 0$, leads to $\sigma = 0$, meaning thus that the source rigidly rotates.

\subsection{Regularity conditions} \label{regcond}

The regularity conditions on the symmetry axis for metric (\ref{metric}) have already been displayed  in \cite{D06}. However, since they will be needed in the following, we recall them briefly here, using our own notations. They imply \cite{S09}
\begin{equation}
\lim_{r \to 0} \frac{g^{\alpha \beta}X_{, \alpha}X_{,\beta}}{4X} = 1, \label{regcond1}
\end{equation}
where $X = g_{\phi \phi}$. Equations (\ref{metric}) and  (\ref{regcond1}) yield
\begin{equation}
\lim_{r \to 0} \frac{e^{-\mu} l'^2}{4l} = 1. \label{regcond2}
\end{equation}
The requirement that $g_{\phi \phi}$ vanishes on the axis implies
\begin{equation}
l \stackrel{0}{=} 0, \label{regcond3}
\end{equation}
where $ \stackrel{0}{=}$ means that the values are taken at $r=0$.

Since, from a physical point of view, there cannot be singularities along the axis, we impose that, at this limit, spacetime tends to flatness; hence we scale the coordinates such that, for $r \rightarrow 0$, the  metric becomes
\begin{equation}
\textrm{d}s^2 = -\textrm{d}t^2 + 2\omega r^2 \textrm{d}t \textrm{d}\phi +\textrm{d}r^2 + \textrm{d}z^2 + r^2 \textrm{d}\phi^2, \label{axismetric}
\end{equation}
from which
\begin{equation}
f  \stackrel{0}{=} 1, \qquad k \stackrel{0}{=} \mu  \stackrel{0}{=} 0, \label{regcond4}
\end{equation}
follow, implying
\begin{equation}
D  \stackrel{0}{=} 0, \label{regcond5}
\end{equation}
and, from (\ref{regcond2}) and (\ref{axismetric}),
\begin{equation}
l'  \stackrel{0}{=} 0. \label{regcond6}
\end{equation}
Then, from the above and the requirement that the Einstein tensor components in  (\ref{G00})-(\ref{G33}) do not diverge, we have
\begin{equation}
f'  \stackrel{0}{=} k' \stackrel{0}{=} k'' - k'\frac{D'}{D} \stackrel{0}{=} 0, \label{regcond7}
\end{equation}
and from  (\ref{regcond6}) and (\ref{regcond7}) we obtain
\begin{equation}
D'  \stackrel{0}{=} k'' \stackrel{0}{=} 0. \label{regcond8}
\end{equation}

\subsection{Junction conditions} \label{junct}

These conditions have also been displayed in \cite{D06} for metric (\ref{metric}). For completeness and also since they will be partially needed further on, we recall them here briefly, in a version adapted to nonrigid rotation and to our own notations. 
 
Outside the fluid cylinder, a vacuum solution to the field equations is needed. Since our system is stationary, the Lewis metric \cite{L32} will be used to represent such an exterior spacetime. And its Weyl class \cite{dS95} is here chosen for junction condition purposes. This metric can be written as
\begin{equation}
\textrm{d}s^2=-F \textrm{d}t^2 + 2 K \textrm{d}t \textrm{d}\phi +\textrm{e}^M (\textrm{d}R^2 +\textrm{d}z^2) + L \textrm{d}\phi^2, \label{Wmetric}
\end{equation}
where
\begin{equation}
F= a R^{1 - n} - a \delta^2 R^{1 + n}, \label{W1}
\end{equation}
\begin{equation}
K = - (1 - ab\delta)\delta R^{1 + n} - ab  R^{1 - n}, \label{W2}
\end{equation}
\begin{equation}
\textrm{e}^M =  R^{(n^2 - 1)/2},  \label{W3}
\end{equation}
\begin{equation}
L = \frac{(1 - ab\delta)^2}{a} R^{1 + n} - ab^2 R^{1 - n}, \label{W4}
\end{equation}
with
\begin{equation}
\delta = \frac{c}{an}, \label{W5}
\end{equation}
where $a, b, c$, and $n$ are real constants. See \cite{D06} for comments about the respective coordinate systems inside and outside the fluid cylinder and \cite{dS95} for more details about the Lewis metric Weyl class.

In accordance with Darmois' junction conditions \cite{D27}, metric (\ref{metric}) and metric (\ref{Wmetric})'s coefficients and their derivatives must be continuous across the $\Sigma$ surface,
\begin{equation}
f  \stackrel{\Sigma}{= }a_1 F, \quad k  \stackrel{\Sigma}{=} a_2 K, \quad \textrm{e}^\mu  \stackrel{\Sigma}{=} a_3 \textrm{e}^M, \quad l  \stackrel{\Sigma}{=} a_4 L, \label{W6}
\end{equation}
\begin{equation}
\frac{f'}{f}  \stackrel{\Sigma}{=} \frac{1}{R} + n \frac{\delta^2 R^n + R^{-n}}{\delta^2 R^{1 + n} - R^{1 - n}}, \label{W7}
\end{equation}
\begin{equation}
\frac{k'}{k}  \stackrel{\Sigma}{=} \frac{1}{R} + n \frac{(1 - ab\delta)\delta R^n - abR^{-n}}{(1 - ab\delta)\delta R^{1 + n} + abR^{1 - n}}, \label{W8}
\end{equation}
\begin{equation}
\mu' \stackrel{\Sigma}{=} \frac{n^2 - 1}{2R},  \label{W9}
\end{equation}
\begin{equation}
\frac{l'}{l}  \stackrel{\Sigma}{=} \frac{1}{R} + n \frac{(1 - ab\delta)^2 R^n + a^2b^2 R^{-n}}{(1 - ab\delta)^2 R^{1 + n} - a^2b^2R^{1 - n}}. \label{W10}
\end{equation}
The first fundamental form continuity imposes  (\ref{W6}) where the $a_1$, $a_2$, $a_3$, and $a_4$ constants can be transformed away by rescaling the coordinates, while (\ref{W7})-(\ref{W10}) are produced by the second fundamental form continuity. Hence, the above equations inserted into (\ref{G11}) imply $P_r \stackrel{\Sigma}{=} 0$,  as expected.

In the low density limit, the $n$ parameter is connected to the Newtonian mass per unit length $\sigma$ of a uniform line mass, as follows \cite{dS95}:
\begin{equation}
\sigma = \frac{1 - n}{4}. \label{sigma}
\end{equation}

Comments about other spacetime properties issuing from the above relations are displayed in \cite{D06}, part of them pertaining exclusively to the rigid rotation case. We do not recall them here since they will not be needed for present purposes.

\subsection{Stress-energy tensor conservation }

Writing the stress-energy tensor conservation is analogous to writing the Bianchi identity
\begin{equation}
T^\beta_{1;\beta} = 0. \label{Bianchi}
\end{equation}
From (\ref{setens}), we have
\begin{eqnarray}
T^{\alpha \beta} = (\rho + P_r) V^\alpha V^\beta + P_r g^{\alpha \beta} + (P_\phi - P_r) K^\alpha K^\beta + (P_z - P_r)S^\alpha S^\beta. \label{setenscontra}
\end{eqnarray}
With $V^\alpha$ given by (\ref{nr4velocity}), and the spacelike vectors $K^\alpha$ and  $S^\alpha$ given, respectively, by (\ref{kalpha}) and (\ref{salpha}), which we insert into (\ref{setenscontra}), using (\ref{metric}) and (\ref{timelike}), Bianchi identity (\ref{Bianchi}) reduces to
\begin{equation}
T^\beta_{1;\beta} = P'_r - (\rho + P_\phi) \Psi + (P_r - P_\phi)\frac{D'}{D} + \frac{1}{2}(P_r - P_z)\mu'  = 0. \label{Bianchi2}
\end{equation}

\subsection{Gravito-electromagnetism} \label{gravelecmagn}

In this section we will study metric (\ref{metric})'s gravito-electromagnetic properties when its Weyl tensor features are coupled to the field equations (\ref{G00})-(\ref{G33}).

\subsubsection{The Weyl tensor} \label{wtensor}

The Weyl tensor nonzero components, $C_{\alpha \beta \gamma \delta}$, valid for both nonrigid and rigid rotation cases, are
\begin{eqnarray}
C_{0101} = \frac{f''}{4} - \frac{f \mu''}{12} - \frac{fD''}{6D} - \frac{f' \mu'}{4} - \frac{f' D'}{4D} + \frac{f \mu' D'}{4D} + \frac{f(f'l' + k'^2)}{6D^2}, \label{C0101}
\end{eqnarray}
\begin{eqnarray}
C_{0202} = - \frac{f''}{4} - \frac{f \mu''}{12} + \frac{fD''}{3D} + \frac{f' \mu'}{4} + \frac{f' D'}{4D} - \frac{f \mu' D'}{4D} - \frac{f(f'l' + k'^2)}{3D^2}, \label{C0202}
\end{eqnarray}
\begin{equation}
C_{0303} = \frac{\textrm{e}^{-\mu}}{6}( - DD'' + D^2\mu'' + f'l' + k'^2), \label{C0303}
\end{equation}
\begin{eqnarray}
C_{0113} = \frac{k''}{4} - \frac{k \mu''}{12} - \frac{kD''}{6D} - \frac{k' \mu'}{4} - \frac{k' D'}{4D} + \frac{k \mu' D'}{4D} + \frac{k(f'l' + k'^2)}{6D^2}, \label{C0113}
\end{eqnarray}
\begin{eqnarray}
C_{0223} = - \frac{k''}{4} - \frac{k \mu''}{12} + \frac{kD''}{3D} + \frac{k' \mu'}{4} + \frac{k' D'}{4D} - \frac{k \mu' D'}{4D} - \frac{k(f'l' + k'^2)}{3D^2}, \label{C0223}
\end{eqnarray}
\begin{equation}
C_{1212} = \frac{\textrm{e}^{\mu}}{6}\left( - \mu'' + \frac{D''}{D} - \frac{f'l' + k'^2}{D^2}\right), \label{C1212}
\end{equation}
\begin{eqnarray}
C_{1313} = - \frac{l''}{4} + \frac{l \mu''}{12} + \frac{lD''}{6D} + \frac{l' \mu'}{4} + \frac{l' D'}{4D} - \frac{l \mu' D'}{4D} - \frac{l(f'l' + k'^2)}{6D^2}, \label{C1313}
\end{eqnarray}
\begin{eqnarray}
C_{2323} = \frac{l''}{4} + \frac{l \mu''}{12} - \frac{lD''}{3D} - \frac{l' \mu'}{4} - \frac{l' D'}{4D} + \frac{l \mu' D'}{4D} + \frac{l(f'l' + k'^2)}{3D^2}. \label{C2323}
\end{eqnarray}
They satisfy the five following relations:
\begin{equation}
C_{1212} = - \frac{\textrm{e}^{2\mu}}{D^2} C_{0303}, \label{rel1}
\end{equation}
\begin{equation}
D^2  (C_{0101} + C_{0202}) + f \textrm{e}^\mu C_{0303} = 0, \label{rel2}
\end{equation}
\begin{equation}
k (C_{0101} + C_{0202}) = f (C_{0113} + C_{0223}), \label{rel3}
\end{equation}
\begin{equation}
l (C_{0101} + C_{0202}) = -f (C_{1313} + C_{2323}), \label{rel4}
\end{equation}
\begin{equation}
l C_{0101} + f C_{2323}  + k (C_{0113} - C_{0223}) = 0. \label{rel5}
\end{equation}
There are thus only three independent Weyl tensor components .

\subsubsection{Conformal flatness} \label{cf1}

As it is well-known, any spacetime whose Weyl tensor vanishes is conformally flat. In this section, we show that, when metric (\ref{metric}) obtains, a null Weyl tensor is incompatible with the regularity conditions on the symmetry axis. As a consequence, conformal flatness is not allowed for the corresponding spacetimes.

Assuming $C_{\alpha\beta\gamma\delta}=0$ for any $\alpha$, $\beta$, $\gamma$, and $\delta$, we should have, in particular, $C_{0303}=0$ and $C_{0223}=0$, which, using (\ref{C0303}) and  (\ref{C0223}), give
\begin{equation}
\frac{D'}{D} = \frac{(k\mu' - k')'}{k\mu' - k'}, \label{C1=0}
\end{equation}
which can be integrated as
\begin{equation}
c_1(k\mu' - k') = D, \label{partint4}
\end{equation}
where $c_1$is an integration constant.

An analogous reasoning conducted with $C_{0303}=0$ and $C_{1313}=0$, using (\ref{C0303}) and (\ref{C1313}), yields
\begin{equation}
\frac{D'}{D} = \frac{(l\mu' - l')'}{l\mu' - l'}, \label{C2=0}
\end{equation}
which can be integrated as
\begin{equation}
c_2(l\mu' - l') = D, \label{partint5}
\end{equation}
where $c_2$ is an integration constant.

Then, from (\ref{partint4}) and (\ref{partint5}), we obtain
\begin{equation}
\mu' = \frac{c_1k' - c_2 l'}{c_1k - c_2l}, \label{partint45}
\end{equation}
which we can integrate as
\begin{equation}
\textrm{e}^\mu = c_3(c_1k - c_2 l), \label{int}
\end{equation}
where $c_3$ is an integration constant.

Now, regularity conditions (\ref{regcond3}) and (\ref{regcond4}) inserted into (\ref{int}) would give
\begin{equation}
1 \stackrel{0}{=} 0, \label{regcondp}
\end{equation}
which is obvious nonsense issuing from the $C_{0303}$, $C_{0223}$ and $C_{1313}$ vanishing assumptions.

We are thus led to the conclusion that, since the three Weyl tensor components $C_{0303}$, $C_{0223}$ and $C_{1313}$ are not allowed to simultaneously vanish, the corresponding spacetimes cannot be conformally flat.

\subsubsection{Purely electric and purely magnetic spacetimes} \label{pem}

Nonconformally flat Weyl tensor electric and magnetic parts, respectively $E_{\alpha \beta}$ and  $H_{\alpha \beta}$, as measured by an observer with 4-velocity $u^\alpha$ (unit timelike congruence), are pointwise defined from the Weyl tensor $C_{\alpha\beta\gamma\delta}$ and its dual $ \tilde{C}_{\alpha \beta \gamma \delta}$ by contraction with the 4-velocity vector $u^\alpha$ as
\begin{equation}
E_{\alpha \beta} = C_{\alpha \gamma \beta \delta}u^\gamma u^\delta, \label{elecdef}
\end{equation}
\begin{eqnarray}
H_{\alpha \beta} &=& \tilde{C}_{\alpha \gamma \beta \delta}u^\gamma u^\delta = \frac{1}{2} \epsilon_{\alpha \gamma \epsilon \delta} C^{\epsilon \delta}_{\beta \rho} u^\gamma u^\rho, \nonumber \\
\epsilon_{\alpha \beta \gamma \delta} &\equiv& \sqrt{-g} \eta_{\alpha \beta \gamma \delta} , \label{magndef}
\end{eqnarray}
where $\eta_{\alpha \beta \gamma \delta} = +1$ or $-1$ for $\alpha$, $\beta$, $\gamma$, $\delta,$ in even or odd order, respectively, and $0$ otherwise.
As  the Weyl tensor itself, its electric and magnetic parts are traceless symmetric tensors, and they determine it entirely.

Spacetimes for which the Weyl tensor magnetic part vanishes are called purely electric while those for which the electric part vanishes are named purely magnetic. Although, while considering (\ref{elecdef}) and (\ref{magndef}), one could presume both the purely electric and the purely magnetic properties to be dependent on the $u^\alpha$ choice, this is actually not the case. If one such property holds, then $u^{\alpha}$ is a Weyl principal vector. Moreover, for Petrov type I spacetimes such as those we will study below, $u^{\alpha}$ is uniquely determined (up to sign) by the Weyl tensor components, $C_{\alpha\beta\gamma\delta}$ \cite{W06b,A92}.

We will show below that stationary nonrigidly rotating cylindrical anisotropic fluid interior solutions can exhibit for some of them a purely electric Weyl tensor and for others a purely magnetic one, provided their metric functions satisfy particular constraints. For this and Petrov classification purposes, we will use the properties of the Weyl tensor complex invariants \cite{M90,P86}. We first define
\begin{equation}
Q_{\alpha \beta} = E_{\alpha \beta} + i H_{\alpha \beta}. \label{Q}
\end{equation}
The quadratic, $I$, cubic, $J$, and zero-dimensional, $M$, invariants, can then be written as \cite{M53,W06a,M94}
\begin{eqnarray}
I = Q^\alpha_\beta Q ^\beta_\alpha = \lambda^2_1 +  \lambda^2_2 +  \lambda^2_3 = -2(\lambda_1\lambda_2 + \lambda_2\lambda_3 + \lambda_3\lambda_1) = E^{\alpha}_{\beta}E^{\beta}_{\alpha} - H^{\alpha}_{\beta}H^{\beta}_{\alpha} + 2i E^{\alpha}_{\beta}H^{\beta}_{\alpha}, \label{I}
\end{eqnarray}
\begin{eqnarray}
J = Q^\alpha_\beta Q^\beta_\gamma Q^\gamma_\alpha =  \lambda^3_1 +  \lambda^3_2 +  \lambda^3_3 = 3 \lambda_1\lambda_2\lambda_3 =  E^{\alpha}_{\beta}E^{\beta}_{\gamma} E^{\gamma}_{\alpha} - i H^{\alpha}_{\beta}H^{\beta}_{\gamma} H^{\gamma}_{\alpha} + 3i  E^{\alpha}_{\beta}(E^{\beta}_{\gamma} + i H^{\beta}_{\gamma}) H^{\gamma}_{\alpha}, \label{J}
\end{eqnarray}
\begin{equation}
M = \frac{2(\lambda_1 -  \lambda_2)^2 (\lambda_2 -  \lambda_3)^2 (\lambda_3 -  \lambda_1)^2}{9 \lambda_1^2\lambda_2^2\lambda_3^2} = \frac{I^3}{J^2} - 6, \label{M}
\end{equation}
where the $\lambda_i$ solutions of equation
\begin{equation}
\lambda^3 -\frac{1}{2} \lambda I - \frac{1}{3}J = 0, \label{eigen}
\end{equation}
are $Q_{\alpha \beta}$'s eigenvalues. A Petrov type $I$ or $D$ Weyl tensor is purely electric (magnetic) iff every $\lambda_i$ is real (imaginary), or, according to \cite{M94}, iff $I$ is real positive (negative) and $M$ is real non-negative or infinite.

The spacetimes considered here have been shown in Sec. \ref{cf1} to be nonconformally flat. We can therefore apply the above rule from Ref. \cite{M94} that implies the Weyl tensor complex invariants to exhibit their properties of interest. Since these invariants are observer's unit velocity 4-vector independent, we choose, for calculation convenience and without loss of generality, a corotating with the fluid observer with unit timelike 4-velocity of the kind
\begin{equation}
u^{\alpha} = v_o \delta^{\alpha}_0, \label{vobs}
\end{equation}
which moreover will be shown below to be actually a Weyl principal vector for the studied solution classes. This 4-velocity, being unit and timelike, thus obeys
\begin{equation}
u^{\alpha}u_{\alpha} = -1, \label{obstl}
\end{equation}
which gives
\begin{equation}
fv^2_o - 1 = 0. \label{tlobs}
\end{equation}

Inserting (\ref{vobs}) into (\ref{elecdef}) and using (\ref{tlobs}) with metric (\ref{metric})'s contravariant coefficients, we obtain the following nonzero Weyl tensor electric part components as measured by such an observer:
\begin{equation}
E_{11} = \frac{C_{0101}}{f}, \label{obsE11}
\end{equation}
\begin{equation}
E_{22} =  \frac{C_{0202}}{f}, \label{obsE22}
\end{equation}
\begin{equation}
E_{33} =  \frac{C_{0303}}{f}, \label{obsE33}
\end{equation}
which are not independent, since, by (\ref{rel2}) virtue, we have
\begin{equation}
D^2 E_{11} + D^2 E_{22} + f \textrm{e}^\mu E_{33} = 0. \label{obsErel}
\end{equation}
Therefore, for this observer's congruence, the Weyl tensor electric part exhibits two independent components.

Then, inserting (\ref{vobs}) into (\ref{magndef}), and using (\ref{tlobs}) and metric (\ref{metric})'s contravariant coefficients, we obtain the only nonzero Weyl tensor magnetic part component as
\begin{equation}
H_{12} =  H_{21} = \frac{1}{D}\left(C_{0223} - \frac{k}{f}C_{0202}\right). \label{obsH12}
\end{equation}

Inserting (\ref{obsE11})-(\ref{obsH12}) into (\ref{I}) and (\ref{J}), while using metric (\ref{metric})'s coefficients, we obtain
\begin{equation}
I = \frac{\textrm{e}^{-2 \mu}}{f^2}\left[ (C_{0101})^2 + (C_{0202})^2 \right] + \frac{(C_{0303})^2}{D^4} - 2 \textrm{e}^{-2 \mu} (H_{12})^2, \label{Ia}
\end{equation}
\begin{equation}
J = \frac{3 \textrm{e}^{-2 \mu}}{f^2 D^2} C_{0303} \left[ C_{0101}C_{0202} + f^2 ( H_{12})^2 \right]. \label{Ja}
\end{equation}
Now, inserting  (\ref{C0202}) and (\ref{C0223}) into  (\ref{obsH12}) gives
\begin{equation}
H_{12} = \frac{\textrm{e}^{\mu}}{4f} \left[ \textrm{e}^{- \mu} \left(\frac{kf' - fk'}{D}\right) \right]'. \label{H12a}
\end{equation}
Inserting (\ref{H12a}) into (\ref{Ia}) and using (\ref{rel2}), $I$ becomes
\begin{eqnarray}
I = \frac{2 \textrm{e}^{-2 \mu}}{f^2} \left[ (C_{0101})^2 + (C_{0202})^2 + C_{0101} C_{0202} \right] - \frac{1}{8 f^2}\left\{\left[\textrm{e}^{- \mu} \left(\frac{kf' - fk'}{D}\right)\right]'\right\}^2. \label{Ib}
\end{eqnarray}
We thus see that both $I$ and $J$ are real, since the metric functions considered here are themselves real.

In the following, we will analyze particular cases which share the same properties: i. e., they all verify $J=0$ and $I \neq 0$. Recalling that any purely electric or magnetic spacetime is Petrov type $I$, we note that, in the extended Petrov classification by Arianrhod and McIntosh \cite{M90,A92}, $\{J=0$, $I \neq 0 \}$ corresponds to $I(M^{\infty})$, with one of the $\lambda_i$ being identically zero. Of course, in this case, $M$ goes to infinity and the rule of McIntosh {\it et al.} \cite{M94} applies. Therefore, we will use this rule to determine whether the corresponding Weyl tensor is purely electric (magnetic), which is the case iff $M$ is real non-negative or infinite and $I$ is real positive (negative). Our task will thus be to determine the $I$ sign. 

From (\ref{Ja}), there are two cases for which $J$ vanishes: either $C_{0303}=0$ or $C_{0101}C_{0202} + f^2 ( H_{12})^2=0$. We will examine below both possibilities.

\hfill

We begin with studying the case for which
\begin{equation}
C_{0303}=0, \label{vanC0303a}
\end{equation}
which, using  (\ref{C0303}), gives
\begin{equation}
\frac{D''}{D} = \mu'' + \frac{f'l' + k'^2}{D^2}. \label{vanC0303b}
\end{equation}
Inserting (\ref{vanC0303a}) into (\ref{Ia}), using  (\ref{H12a}), $I$ becomes
\begin{equation}
I =  \frac{\textrm{e}^{-2 \mu}}{f^2}\left[ (C_{0101})^2 + (C_{0202})^2 \right] - \frac{1}{8 f^2}\left\{\left[\textrm{e}^{- \mu} \left(\frac{kf' - fk'}{D}\right)\right]'\right\}^2. \label{Ic}
\end{equation}
We see from (\ref{Ic}) that $I$ is the sum of a positively defined term and of a negatively defined one. Without loss of generality, we can therefore assume 
 $I \neq 0$, since, given a solution of the field equations (\ref{G00})-(\ref{G33}) verifying also (\ref{vanC0303b}), the $I$ invariant may possibly vanish only for an {\it a priori} countable $r$ value set or it should be identically nonzero. Hence, the rule of McIntosh {\it et al.} \cite{M94} applies. Notice that, with $I \neq 0$, once $C_{0303}$ is fixed by (\ref{vanC0303a}), two independent Weyl tensor nonzero components still remain.

\hfill

Now, using (\ref{obsH12}) into $\left[ C_{0101}C_{0202} + f^2 ( H_{21})^2 \right] = 0$, this other relation implying $J = 0$ can be written
\begin{equation}
D^2 C_{0101}C_{0202} + \left(fC_{0223} - k C_{0202}\right)^2 = 0. \label{vanJb}
\end{equation}
Then, inserting  (\ref{C0202}) and  (\ref{C0223}) into  (\ref{vanJb}), we obtain
\begin{eqnarray}
D^2 C_{0101}C_{0202} + \frac{1}{16}\left[(fk' - kf')\left(\mu' + \frac{D'}{D} \right) - (fk' - kf')' \right]^2 = 0. \label{vanJc}
\end{eqnarray}
Recall that every spacetime whose metric functions verify (\ref{Ic}) or (\ref{vanJc}), implying $J=0$, is therefore Petrov type $I(M^\infty)$.

Now, we will analyze the purely electric and purely magnetic subsamples.

\hfill

1. {\it Purely electric spacetimes} -- Recall that any $J=0$ Weyl tensor is purely electric provided its $I$ invariant is positive. 

Beginning with the (\ref{Ic}) case, which we will refer to as the first $J=0$ case, we see that a sufficient, while not necessary, condition is 
\begin{equation}
\left[\textrm{e}^{- \mu} \left(\frac{kf' - fk'}{D}\right)\right]' = 0, \label{cond11a}
\end{equation}
which can be integrated as
\begin{equation}
kf' - fk' = c_4 \textrm{e}^\mu D, \label{cond11b}
\end{equation}
$c_4$ being an integration constant. It can easily be checked that above Eq. (\ref{cond11b}) actually verifies the regularity conditions displayed in Section \ref{regcond}. Now, from  (\ref{obsH12}) and  (\ref{Ia}), we see  that (\ref{cond11a}) imposes, on the Weyl tensor components, a new constraint that reads
\begin{equation}
f C_{0223} - k C_{0202} = 0. \label{cond11c}
\end{equation}
However, one nonzero independent component still remains since we have seven relations, (\ref{rel1})-(\ref{rel5}),  (\ref{vanC0303a}), and (\ref{cond11c}), for eight nonzero $C_{\alpha \beta \gamma \delta}$. The corresponding spacetimes are thus nonconformally flat.

Then, we simplify once more the field equations by inserting (\ref{cond11b}) into (\ref{partint2}), which gives
\begin{equation}
\mu' = \frac{2 \kappa}{c_4} D (\rho + P_\phi) (k\Omega^2 - fv\Omega). \label{muprime}
\end{equation}
Notice that in the nonrigid rotation case where $\Omega \neq 0$, (\ref{muprime}) implies $\mu' \neq 0$, thus $\sigma \neq 0$, and hence a nonflat exterior (see Sec. \ref{elecr} for an inverse detailed demonstration). Therefore the interior spacetime is not bound to be static.

\hfill

Now, we consider the second $J=0$ case and insert (\ref{cond11b}) into (\ref{vanJc}). We obtain
\begin{equation}
C_{0101}C_{0202} = 0, \label{cond21a}
\end{equation}
which has two solutions

\hfill

$C_{0101} = 0$

where we insert (\ref{C0101}) that gives
\begin{equation}
\frac{f''}{2} - \frac{f \mu''}{6} - \frac{f D''}{3D} - \frac{f' \mu'}{2} - \frac{f' D'}{2D} + \frac{f \mu' D'}{2D} + \frac{f(f'l' + k'^2)}{3 D^2} =0.  \label{cond211}
\end{equation}

\hfill

$C_{0202} = 0$

where we insert (\ref{C0202}) that gives
\begin{equation}
- \frac{f''}{4} - \frac{f \mu''}{12} + \frac{f D''}{3D} + \frac{f' \mu'}{4} + \frac{f' D'}{4D} - \frac{f \mu' D'}{4D} - \frac{f(f'l' + k'^2)}{3 D^2} =0.  \label{cond212}
\end{equation}

\hfill

We have thus implicitly defined three classes of {\it purely electric} interior spacetimes sourced by a stationary nonrigidly rotating anisotropic fluid whose metric functions are solutions of the seven differential equations (\ref{G00})-(\ref{G33}), (\ref{vanC0303b}), and either, respectively, (\ref{cond11b}), (\ref{cond211}), or (\ref{cond212}), and of the timelike condition (\ref{timelike}). Notice that, in the first $J=0$ case, four among these equations can be replaced by (\ref{partint1})-(\ref{partint3}) and (\ref{muprime}) which are partially integrated equations, i. e., interesting simplifications for future analytic or numerical resolutions. This eight equation set exhibits ten unknown $r$ functions: the four metric functions and the six fluid parameters which have thus to be determined by an appropriate equation of state choice, none of these eight equations implying necessary staticity for the solutions. These considerations lead us to conjecture that such purely electric spacetimes do actually exist. However, a large number of mathematically {\it and physically} interesting purely electric solutions can be found in the literature, e. g., all the static spacetimes \cite{S09}. This is at variance with the purely magnetic type of which scarce examples are known \cite{W06b,L07}. A large tractable part of such solutions sourced by the here studied fluid and verifying $J = 0$, are analyzed in below subsection.

\hfill

2. {\it Purely magnetic spacetimes} -- Recall that, from the rule of McIntosh {\it et al.} \cite{M94}, any $\{J=0, I<0\}$ spacetime is purely magnetic. 

Now, the vanishing of the first term in Eq. (\ref{Ic}) is a sufficient, while not necessary, condition for $I$ to be negative, and this implies
\begin{equation}
C_{0101} = C_{0202} = 0, \label{vanC0102a}
\end{equation}
together with (\ref{vanC0303a}) which still obtains. Hence, in this case, three Weyl tensor components vanish. However, these components are linked by a linear constraint through (\ref{rel2}). We are thus left, here also, with one nonzero independent Weyl tensor component, and therefore spacetimes of this class are nonconformally flat .

Considering the first $J=0$ case, we can write $C_{0101} = C_{0303} = 0$ into which we insert (\ref{C0101}) and (\ref{C0303}), to obtain
\begin{equation}
\frac{D'}{D} = \frac{(f \mu' - f')'}{f \mu' - f'}, \label{cond12a}
\end{equation}
which can be integrated as
\begin{equation}
c_5 (f \mu' - f') = D, \label{cond12b}
\end{equation}
where we insert (\ref{regcond4}),  (\ref{regcond5}), and (\ref{regcond7}) to obtain a new regularity condition on the $z$ axis, which we write as
\begin{equation}
\mu' \stackrel{0}{=} 0. \label{cond12c}
\end{equation}
Then, removing all the second derivatives from (\ref{G00}),  (\ref{G11}),  (\ref{G22}),  $C_{0101} = 0$, and $C_{0303} = 0$ into both of which we insert (\ref{C0101}) and (\ref{C0303}), we obtain
\begin{equation}
\mu' \textrm{e}^\mu \left(f' - 4 \frac{f D'}{D} \right) = 2 \kappa \left[(\rho - 2 P_r + P_z) f  + (\rho + P_{\phi}) D^2 \Omega^2 \right]. \label{cond12d}
\end{equation}

We have therefore implicitly defined a class of {\it purely magnetic} interior spacetimes sourced by a stationary nonrigidly rotating anisotropic fluid whose metric functions are solutions of the seven differential equations (\ref{G00})-(\ref{G33}), (\ref{vanC0303b}), and (\ref{cond12b}) and of the timelike condition (\ref{timelike}). Four among these equations can be replaced by (\ref{partint1})-(\ref{partint3}) and (\ref{cond12d}) which are partially integrated equations. As for the purely electric case, this eight equation set exhibits ten unknown $r$ functions. Analogous considerations lead us therefore to conjecture that such purely magnetic spacetimes actually exist. However, an extremely small number of mathematically, and moreover {\it physically}, interesting purely magnetic solutions can be found in the literature \cite{W06b,L07}. Our strongly based conjecture is therefore significant progress toward the purely magnetic property understanding.

\hfill

Now, we display the analysis of the second $J=0$ case.

From (\ref{rel2}), (\ref{vanC0102a}) implies $C_{0303} = 0$, and (\ref{vanJb}) gives
\begin{equation}
f C_{0223} - kC_{0202} = 0, \label{cond22}
\end{equation}
from which we obtain, with (\ref{vanC0102a}), $C_{0223} = 0$. Now, while $C_{0101}$, $C_{0202}$, and $C_{0303}$ are not independent from one another by virtue of (\ref{rel2}), (\ref{rel3})-(\ref{rel5}) consideration shows that $C_{0223}$ is actually independent of the other two. Three independent Weyl tensor component vanishing implies the corresponding spacetimes have got a null Weyl tensor. They are therefore conformally flat and, in this case, the rule of McIntosh {\it et al.} \cite{M94} does not apply.

\hfill

In Sec. \ref{pem}, we have therefore exhibited the existence, and partially characterized, three classes of purely electric and one class of purely magnetic  interior, i. e., {\it nonvacuum}, spacetimes, sourced by a nonrigidly rotating stationary cylindrical anisotropic fluid.

\section{Rigidly rotating fluid} \label{rr}

Now we come to the rigid rotation case which has been already considered in \cite{D06} but needs the study of some essential points that have not been analyzed there and a few clarifications.

In the following, we will adapt the analysis made in Sec. \ref{nrrf} for the nonrigid rotation case to the rigid rotation one. We have shown, in Sec. \ref{tobetitled}, that rigid rotation occurs when $\Omega = 0$. From (\ref{timelike}), this implies $fv^2 = 1$. We will thus obtain the relations and expressions pertaining to rigid rotation by inserting
\begin{equation}
\Omega = 0, \quad fv^2 = 1, \label{rigidconstr}
\end{equation}
into the equations written for the nonrigid case, provided they are not issued from dividing or multiplying by $\Omega$, i.e., zero.

\subsection{Field equations} \label{rfe}

Inserting (\ref{rigidconstr}) into (\ref{G00})-(\ref{G33}), we obtain \cite{D06}
\begin{eqnarray}
G_{00} = \frac{\textrm{e}^{-\mu}}{2} \left[-f\mu'' - 2f\frac{D''}{D} + f'' - f'\frac{D'}{D} + \frac{3f(f'l' + k'^2)}{2D^2}\right] = \kappa \rho f, \label{G00r}
\end{eqnarray}
\begin{eqnarray}
G_{03} =  \frac{\textrm{e}^{-\mu}}{2} \left[k\mu'' + 2 k \frac{D''}{D} -k'' + k'\frac{D'}{D} - \frac{3k(f'l' + k'^2)}{2D^2}\right] = - \kappa \rho k, \label{G03r}
\end{eqnarray}
\begin{equation}
G_{11} = \frac{\mu' D'}{2D} + \frac{f'l' + k'^2}{4D^2} = \kappa P_r \textrm{e}^\mu, \label{G11r}
\end{equation}
\begin{equation}
G_{22} = \frac{D''}{D} -\frac{\mu' D'}{2D} - \frac{f'l' + k'^2}{4D^2} = \kappa P_z \textrm{e}^\mu, \label{G22r}
\end{equation}
\begin{eqnarray}
G_{33} =  \frac{\textrm{e}^{-\mu}}{2} \left[l\mu'' + 2l\frac{D''}{D} - l'' + l'\frac{D'}{D} - \frac{3l(f'l' + k'^2)}{2D^2}\right] = \frac{\kappa}{f} (\rho k^2 + P_\phi D^2). \label{G33r}
\end{eqnarray}
With $\Omega = 0$, Eq. (\ref{partint2}) becomes
\begin{equation}
\left(\frac{kf' - fk'}{D}\right)' = 0, \label{partint6}
\end{equation}
which we can integrate as in \cite{D06}
\begin{equation}
kf' - fk' = c_6 D, \label{partint7}
\end{equation}
where $c_6$ is an integration constant. It is easy to verify that inserting the above rigid constraints (\ref{rigidconstr}) into the remaining equation of (\ref{partint}) gives the same result as (\ref{partint7}).

\subsection{Hydrodynamical scalars, vectors, and tensors} \label{hydrdynr}

The fluid 4-velocity, satisfying conditions (\ref{fourvec}), can be chosen in the rigid rotation case by setting $\Omega = 0$ in (\ref{nr4velocity}) which gives
\begin{equation}
V^\alpha = v \delta^\alpha_0. \label{r4velocity}
\end{equation}
The two spacelike 4-vectors (\ref{kalpha}) and (\ref{salpha}) chosen to define the stress-energy tensor and verifying conditions (\ref{fourvec}) become 
\begin{equation}
K^\alpha = -\frac{v}{D}\left(k\delta^\alpha_0 + f \delta^\alpha_3\right), \label{rkalpha}
\end{equation}
\begin{equation}
S^\alpha = \textrm{e}^{-\frac{\mu}{2}}\delta^\alpha_2, \label{rsalpha}
\end{equation}
respectively.

The timelike 4-vector $V_\alpha$ can be as well invariantly decomposed into three independent parts through the genuine tensor $V_{\alpha;\beta}$ as in (\ref{decomp}). Now, $\Psi$, as defined by (\ref{psidef}), becomes, with the rigid rotation constraints (\ref{rigidconstr}),
\begin{equation}
\Psi = fvv' = -\frac{1}{2}f' v^2 = - \frac{f'}{2f}, \label{rpsi}
\end{equation}
which we insert into (\ref{dotV1}) to obtain the acceleration vector only nonzero component as
\begin{equation}
\dot{V}_1 = \frac{f'}{2f}. \label{rdotV1}
\end{equation}
Substituting $\Omega = 0$ and (\ref{rpsi}) into (\ref{omega01})-(\ref{sigma13}), we obtain, respectively,
\begin{equation}
2 \omega_{01} =  - (f'v + 2 fv'), \label{romega01}
\end{equation}
\begin{equation}
2 \omega_{13} = - (k'v + 2kv'), \label{romega13}
\end{equation}
\begin{equation}
\sigma_{01} = 0, \label{rsigma01}
\end{equation}
\begin{equation}
\sigma_{13} = 0, \label{rsigma13}
\end{equation}
which confirms that rigid rotation is actually shear-free.

Now, inserting $\Omega = 0$ and (\ref{rpsi}) into (\ref{modaccel}) and (\ref{omega2}), gives, respectively, the acceleration vector modulus
\begin{equation}
\dot{V}^\alpha \dot{V}_\alpha = \textrm{e}^{-\mu} f v'^2, \label{rmodaccel}
\end{equation}
and the rotation scalar, $\omega$, from
\begin{equation}
\omega^2 = \frac{1}{4f^2  \textrm{e}^\mu D^2}(kf' - fk')^2, \label{romega2}
\end{equation}
which becomes, while inserting (\ref{partint7}),
\begin{equation}
\omega^2 = \frac{c_6^2}{4f^2  \textrm{e}^\mu}.
\end{equation}
We stress that this rotation scalar depends only on the metric functions $f$ and $\textrm{e}^\mu$ and that it cannot vanish, since $c_6 = 0$ would imply $k=0$, and hence the cylindrical static Levi-Civita {\it vacuum} solution which does not pertain to the {\it interior} solution class we are considering here.

\subsection{Regularity conditions} \label{rregcond}

The regularity conditions on the axis which have been displayed in Sec. \ref{regcond} above still obtain in the rigid case since they only depend on the metric functions and not on the fluid parameters. Hence, we will refer to them in the following without any further rotation type specification. 

\subsection{Junction conditions} \label{rjunct}

As the regularity conditions, the junction conditions which have been displayed in Sec. \ref{junct} are still valid in the rigid rotation case. However, using the fluid properties pertaining solely to rigid rotation, we can derive a couple of new interesting results. Inserting  (\ref{W1}),  (\ref{W2}),  (\ref{W4}),  (\ref{W5}), (\ref{W7}) and (\ref{W8}) into (\ref{partint7}), we obtain $c_6 = \pm 2 c$. However, since the sign can be absorbed into the constant definitions, we write \cite{D06}
\begin{equation}
c_6 = 2 c. \label{c}
\end{equation}

\subsection{Stress-energy tensor conservation }

Inserting (\ref{rpsi}) into (\ref{Bianchi2}), we obtain the stress-energy tensor conservation equation, i.e., Bianchi identity, which reads
\begin{equation}
T^\beta_{1;\beta} = P'_r + \frac{1}{2} (\rho + P_\phi)\frac{f'}{f} + (P_r - P_\phi)\frac{D'}{D} + \frac{1}{2}(P_r - P_z)\mu'  = 0, \label{Bianchi3}
\end{equation}
analagous to Eq. (12) of Ref. \cite{D06} where a typo correction should be made \footnote{$P_r$ has to be replaced by $P_\phi$ there in the right hand side's second term.}.

\subsection{The Weyl tensor} \label{rgravelecmagn}

The Weyl tensor nonzero components are the same in the rigid rotation case as in the nonrigid one since they depend only on the metric functions. They are thus given by (\ref{C0101})-(\ref{C2323}) and obey the (\ref{rel1})-(\ref{rel5}) constraints which imply there are only three independent Weyl tensor components. The same arguments as displayed in Sec. \ref{cf1} imply that the corresponding spacetimes are also nonconformally flat.

Now, we will study these solutions' purely electric and purely magnetic properties. Owing to the above statements, the complex Weyl tensor invariants are the same as in the nonrigid rotation case. Moreover, since the considered spacetimes are nonconformally flat and since we will still consider only the $J=0$ case, they are of Petrov type $I(M^\infty)$ and the rule of \cite{M94} applies.

Therefore, the differences we will encounter here with respect to the nonrigid rotation problem will only be issued from the field equations and related results, (\ref{partint7}) in particular, the main ones being encompassed into the relations of (\ref{rigidconstr}).

\subsubsection{Purely electric spacetimes} \label{elecr}

For the rigid rotation case, $\Omega = 0$ inserted into the first $J=0$ case equation, (\ref{muprime}), gives $\mu' = 0$, hence $\mu =$ const., which becomes, with the regularity condition (\ref{regcond4}), $\mu = 0$. Now, using matching condition (\ref{W9}), we obtain $n=1$ which gives, from (\ref{sigma}), $\sigma = 0$, in the low density limit. The $\sigma$ Newtonian mass per unit length vanishing produces a flat, though non-Minkowskian, exterior spacetime, sourced by spinning stationary strings \cite{dS95}.

For the second $J=0$ case, the sufficient, but not necessary, condition for $I$ to be positive, i. e., (\ref{cond11a}) has been shown to imply (\ref{cond11b}), which, divided by (\ref{partint7}), gives
\begin{equation}
 \textrm{e}^\mu = \frac{c_6}{c_4}, \label{vanJe}
\end{equation}
that implies $\mu' = 0$, which results, as above, in $\mu = 0$, hence $\sigma = 0$.

We are thus led to {\it conjecture} that any purely electric interior solution for spacetimes sourced by {\it rigidly rotating} stationary cylindrical anisotropic fluid are {\it flat}. However, we have given an actual {\it proof} of this statement only for the studied subclasses which do not exhaust the $I>0$ condition.

This is at variance with the statement that purely electric solutions should be necessarily static, as concluded improperly in \cite{D06} \footnote{The reasoning in \cite{D06} is not rigourous actually since $kf' - fk' = 0$ does not result in $k = 0$ [their regularity conditions (22) merely give $ 0 - 0  \stackrel{0}{=} 0$], but instead in $D = 0$, from their equation (14), our (\ref{partint7}). This would imply, owing to the metric signature which imposes the metric functions $f$, $l$, and $k$ to be positively defined, $f = l = k =0$, hence an absurd result for the metric. Therefore, the only solution to their Eq. (72) is $\gamma' = 0$, i.e., $\mu' = 0$ in our notations.}. Our own method, using invariants which are the observer's unit velocity 4-vector independent, gives instead proper general results which we have displayed here while stressing their application limits.

\subsubsection{\it Purely magnetic spacetimes}

The only $J=0$ case to be considered here is the first one, since the second one leads to conformally flat spacetimes as shown in Sec. \ref{pem}. The  sufficient condition for $I$ to be negative is still (\ref{vanC0102a}) as in Sec. \ref{pem} § 2. We have seen that this equation can be integrated as (\ref{cond12b}), which therefore obtains also here. Inserting $\Omega = 0$ into (\ref{cond12d}) gives
\begin{equation}
\mu' \textrm{e}^\mu \left(\frac{f'}{f} - 4 \frac{D'}{D} \right) = 2 \kappa (\rho - 2 P_r + P_z). \label{cond12e}
\end{equation}

Such purely magnetic spacetimes are therefore solutions of the five (\ref{G00r})-(\ref{G33r}) field equations, where one among these equations can be replaced by the partially integrated (\ref{partint7}), plus the two (\ref{cond12b}) and (\ref{cond12e}) constraint equations, for four metric functions and four fluid parameters, which means seven equations for eight unknown $r$ functions. Provided an eos is added, this set might be solvable. However, this eos should not be too tight, while, if so, the necessary number of degrees of freedom is overcome and the problem becomes overdetermined, thus, generally, unsolvable. To exemplify this statement we display in Sec. \ref{example} the analysis of a particular eos leading to an exact purely magnetic solution class and, in Appendix A, a special subcase of this eos already considered in \cite{D06} and that we show to be ruled out as a (\ref{G00r})-(\ref{G33r}), (\ref{cond12b}), (\ref{cond12e}) solution.

Moreover, recall we have justified in Sec. \ref{hydrdynr} that these rigidly rotating fluids exhibit nonvanishing rotation scalars as is demanded for a rotating fluid and vanishing shears known as pertaining to rigid rotation. Our results show therefore that in the cylindrically symmetric case, even shear-free fluid motion can source purely magnetic spacetimes, provided rotation should be involved.

\section{Interior spacetimes sourced by stationary cylindrical anisotropic rigidly rotating fluids: purely magnetic exact solutions} \label{example}

Here, we consider the rigid rotation case for a peculiar fluid eos chosen as
\begin{equation}
P_r = P_\phi = 0. \label{eosA}
\end{equation}
Such an equation of state could describe, e. g., a rigidly rotating astrophysical object issuing jets in its symmetry axis direction.
With (\ref{eosA}) inserted, the rigidly rotating fluid field equations  (\ref{G00r})-(\ref{G33r}) become
\begin{equation}
-f\mu'' - 2f\frac{D''}{D} + f'' - f'\frac{D'}{D} + \frac{3f(f'l' + k'^2)}{2D^2} = 2 \kappa \rho f \textrm{e}^{\mu}, \label{G00rA}
\end{equation}
\begin{equation}
k\mu'' + 2 k \frac{D''}{D} -k'' + k'\frac{D'}{D} - \frac{3k(f'l' + k'^2)}{2D^2} = - 2 \kappa \rho k \textrm{e}^{\mu}, \label{G03rA}
\end{equation}
\begin{equation}
2 \mu' D D' + f'l' + k'^2 = 0, \label{G11rA}
\end{equation}
\begin{equation}
\frac{D''}{D} -\frac{\mu' D'}{2D} - \frac{f'l' + k'^2}{4D^2} = \kappa P_z \textrm{e}^\mu, \label{G22rA}
\end{equation}
\begin{equation}
l\mu'' + 2l\frac{D''}{D} - l'' + l'\frac{D'}{D} - \frac{3l(f'l' + k'^2)}{2D^2} = \frac{2 \kappa \rho k^2 \textrm{e}^{\mu}}{f} , \label{G33rA}
\end{equation}
and the (\ref{Bianchi3}) Bianchi identity becomes
\begin{equation}
\mu' = \frac{\rho}{P_z}\frac{f'}{f}. \label{A1}
\end{equation}
The field equations (\ref{G00rA}) and (\ref{G22rA}), give, using (\ref{G11rA}),
\begin{eqnarray}
&&4 \left(1 + \frac{P_z}{\rho} \right) f D D'' + \frac{P_z}{\rho} \left[ 2 f D^2 \mu'' - 2 D^2 f'' \right. \nonumber \\
&& \left. + 2 f' D D' - 3f(f'l' + k'^2) \right] = 0.  \label{5a}
\end{eqnarray}
Now, inserting (\ref{c}) into (\ref{partint7}) we obtain
\begin{equation}
k' = \frac{kf' - 2c D}{f}, \label{5b}
\end{equation}
which we substitute into (\ref{G11rA}) together with (\ref{A1}) such as to obtain, with the help of (\ref{D2}) and its derivative with respect to $r$,
\begin{equation}
2 \left(1 + \frac{P_z}{\rho} \right) ff' D' -  \frac{P_z}{\rho} D (f'^2 - 4c^2) = 0. \label{5c}
\end{equation}
Now, we differentiate (\ref{A1}) with respect to $r$, and obtain
\begin{equation}
\mu'' = \frac{\rho}{P_z} \frac{f''}{f} - \frac{\rho}{P_z} \frac{f'^2}{f^2} + \frac{P_z \rho' - \rho P'_z}{P_z^2} \frac{f'}{f}. \label{5ca}
\end{equation}
Then, we insert (\ref{A1}) into (\ref{G11rA}) which becomes
\begin{equation}
f'l' + k'^2 = -2 \frac{\rho}{P_z} \frac{f'}{f} DD'. \label{5d}
\end{equation}
Now, inserting (\ref{5c}) and (\ref{5d}) into (\ref{5a}) gives
\begin{eqnarray}
2 \left(1 + \frac{P_z}{\rho} \right) f^2 D'' +  \left(1 - \frac{P_z}{\rho} \right)Dff'' - Df'^2 \nonumber \\    
+ \left(\frac{\rho'}{\rho} - \frac{P'_z}{P_z} \right) Dff' + \left(3 + \frac{P_z}{\rho} \right) ff'D' = 0.  \label{5e}
\end{eqnarray}
Then, using (\ref{5c}) and its derivative with respect to $r$ into Eq. (\ref{5e}), we obtain, after some rearrangements,
\begin{eqnarray}
\left(f'^2 + 4c^2 \frac{P_z}{\rho} \right)\left[2 \left(1 + \frac{P_z}{\rho} \right)ff'' + 2\left(\frac{\rho'}{\rho} - \frac{P'_z}{P_z} \right) ff' - \left(2 + \frac{P_z}{\rho} \right)f'^2 + 4c^2  \frac{P_z}{\rho} \right] = 0. \label{5f}
\end{eqnarray}
Notice that (\ref{5f}) proceeds solely from the field equations and from the Bianchi identity which can replace one of them, with the  (\ref{eosA}) eos substituted. This equation has two solutions. They will both be considered in turn in the following.

\subsection{First solution of (\ref{5f})} \label{fs1}

This first solution obtains for
\begin{equation}
f'^2 + 4c^2 \frac{P_z}{\rho} = 0, \label{5g}
\end{equation}
which can be written as
\begin{equation}
f' = \pm 2c \sqrt{\frac{- P_z}{\rho}}. \label{5h}
\end{equation}
The above equation implies of course $P_z<0$. While a negative pressure does not describe a standard fluid, it has been considered anyhow for some applications, e. g., in cosmology where dark energy acts as a negative pressure. Therefore, we will study this solution for completeness.

Equation (\ref{A1}), with (\ref{5h}) inserted, becomes
\begin{equation}
\mu' = \pm \frac{2c}{f} \sqrt{\frac{- \rho}{P_z}}. \label{5i}
\end{equation}

Now, we want to find under which conditions this general relativity (GR) solution verifies the two purely magnetic constraints (\ref{cond12b}) and (\ref{cond12e}). If both constraints are satisfied, the corresponding spacetime is purely magnetic. If they are not, we can say nothing about the solution gravitomagnetic nature, since (\ref{cond12b}) and (\ref{cond12e}) are sufficient but not necessary conditions for the spacetime to be purely magnetic.

Thus, we insert (\ref{5h}) and (\ref{5i}) into (\ref{cond12b}) and obtain
\begin{equation}
D = \pm 2c c_5 \left(\sqrt{\frac{- \rho}{P_z}} - \sqrt{\frac{- P_z}{\rho}} \right). \label{5j}
\end{equation}
Now, we apply the regularity conditions (\ref{regcond5}) to (\ref{5j}) and (\ref{regcond7}) to (\ref{5g}), keeping in mind that $c \neq 0$ and $c_5 \neq 0$, or otherwise we would have $f'=\mu'=D=0$ for all $r$ from (\ref{5h})-(\ref{5j}). 

We thus obtain
\begin{equation}
 P_z  \stackrel{0}{=} \rho  \stackrel{0}{=} 0, \label{5k}
\end{equation}
which is a constraint on the $\rho(r)$ and $P_z(r)$ functions vanishing at $r=0$.

Equation (\ref{5j}) differentiated with respect to $r$ gives
\begin{equation}
D' = \pm c c_5 \left(\frac{P'_z}{P_z} - \frac{\rho'}{\rho} \right)  \left(\sqrt{\frac{- \rho}{P_z}} + \sqrt{\frac{- P_z}{\rho}} \right). \label{5l}
\end{equation}

We use (\ref{eosA}), (\ref{5h}), (\ref{5i}), (\ref{5j}) and (\ref{5l}) into (\ref{cond12e}) second purely magnetic constraint equation and obtain
\begin{equation}
\textrm{e}^\mu = \frac{\kappa}{2c} \frac{A f^2}{B + C f}, \label{5m}
\end{equation}
where we have defined
\begin{equation}
A(r) = P_z^2(\rho^2 - P_z^2), \label{5n}
\end{equation}
\begin{equation}
B(r) = c P_z^2(\rho - P_z), \label{5o}
\end{equation}
\begin{equation}
C(r) = \pm \sqrt{\frac{- P_z}{\rho}}(P_z \rho' - \rho P'_z)(\rho + P_z). \label{5p}
\end{equation}
Then, we differentiate (\ref{5m}) with respect to $r$ and insert (\ref{5h}) and (\ref{5i}) to obtain the second order in $f$ equation couple, one equation for each plus and minus sign:
\begin{eqnarray}
\pm 2c \left(\sqrt{\frac{- \rho}{P_z}} - 2 \sqrt{\frac{- P_z}{\rho}} \right) B + \left[ \pm 2c \left(\sqrt{\frac{- \rho}{P_z}} - \sqrt{\frac{- P_z}{\rho}} \right) C + \frac{AB' - A'B}{A} \right] f + \left(\frac{AC' - A'C}{A} \right) f^2= 0, \label{5q}
\end{eqnarray}
whose coefficients are functions of $\rho(r)$, of $P_z(r)$, and of their first and second derivatives. Since we require $f$ to be real, this implies a constraint on $\rho(r)$, $P_z(r)$, and their first and second derivatives that reads
\begin{eqnarray}
\Delta = \left[\frac{AB' - A'B}{A}  \pm 2c \left(\sqrt{\frac{- \rho}{P_z}} - \sqrt{\frac{- P_z}{\rho}} \right) C \right] ^2 \mp 8c \left(\sqrt{\frac{- \rho}{P_z}} - 2 \sqrt{\frac{- P_z}{\rho}} \right)\left(\frac{AC' - A'C}{A} \right) B \geq 0. \label{5r}
\end{eqnarray}
If $\rho(r)$ and $P_z(r)$ are such that they verify (\ref{5r}) for all $r$, at least with either the plus or the minus sign, then $f(r)$ follows as a (\ref{5q}) real root:

 - if $\Delta = 0$, we have one $f(r)$ solution for each $\{\rho(r),P_z(r)\}$ couple verifying (\ref{5r}), which can be doubled if both signs plus and minus obtain in (\ref{5r}). This solution is
\begin{equation}
f = - \frac{ \left[\frac{AB' - A'B}{A} \pm 2c \left(\sqrt{\frac{- \rho}{P_z}} - \sqrt{\frac{- P_z}{\rho}} \right) C \right]}{\frac{2(AC' - A'C)}{A}}. \label{5qa}
\end{equation}

- if $\Delta > 0$, we have two $f(r)$ solutions for each $\{\rho(r),P_z(r)\}$ couple verifying (\ref{5r}), which can be doubled if both signs plus and minus obtain in (\ref{5r}). They read
\begin{equation}
f_\epsilon =  \frac{- \left[\frac{AB' - A'B}{A} \pm 2c \left(\sqrt{\frac{- \rho}{P_z}} - \sqrt{\frac{- P_z}{\rho}} \right) C \right]+\epsilon \sqrt{\Delta}}{\frac{2(AC' - A'C)}{A}} , \label{5qb}
\end{equation}
where $\epsilon$ can take the values $\pm1$ independently of the $\pm$ sign coming from (\ref{5h}).

We have thus obtained $f(r)$, which we insert into (\ref{5m}) to obtain $\textrm{e}^\mu(r)$. We already know $D(r)$ as given by (\ref{5j}). Then, inserting  (\ref{5h}) and (\ref{5j}) into (\ref{5b}), we obtain
\begin{equation}
\frac{1}{2c} \sqrt{\frac{-\rho}{P_z}} fk' \mp k  \pm 2cc_5 \left( \frac{\rho}{P_z} -1 \right) = 0, \label{5s}
\end{equation}
which is a first order differential equation for $k(r)$ that, knowing $f(r)$ from (\ref{5qa}) or (\ref{5qb}) depending on whether $\Delta \geq 0$ vanishes or not, we can integrate as
\begin{equation}
k = \textrm{e}^{\int_{r_0}^r - Q(u) \textrm{d}u} \left\{c_7  + \int_{r_0}^r S(v) \left[ -  \textrm{e}^{\int_{r_0}^v Q(u) \textrm{d}u} \right] \textrm{d}v \right\}, \label{5sa}
\end{equation}
where $c_7$ is an integration constant, $r_0$ is a constant integration limit verifying $r_0 < r_{\Sigma}$, and
\begin{equation}
Q(r)  = \mp \frac{2c}{f} \sqrt{\frac{- P_z}{\rho}}, \label{5sb}
\end{equation}
\begin{equation}
S(r)  = \pm \frac{4c^2 c_5}{f}\left(  \sqrt{\frac{- \rho}{P_z}} - \sqrt{\frac{- P_z}{\rho}} \right). \label{5sc}
\end{equation}
It is easy to see, considering (\ref{regcond4}) and (\ref{regcond7}), that  (\ref{5s}) is consistent with the (\ref{5k}) regularity condition.

Knowing $D(r)$, $f(r)$, and $k(r)$ through $\rho(r)$, $P_z(r)$, and derivatives, we use (\ref{D2}) to compute the last $l(r)$ metric function as
\begin{equation}
l = \frac{D^2 - k^2}{f}. \label{5o}
\end{equation}

However, since we have five field and two purely magnetic constraint differential equations for six free functions -- $f(r)$, $k(r)$, $\textrm{e}^\mu(r)$, $l(r)$, $\rho(r)$ and $P_z(r)$ -- we should be able to derive, from one or from a combination of the seven differential equations at hand, constraint equations for $\rho(r)$, $P_z(r)$, and possibly their first and second derivatives. For this purpose, we insert (\ref{G11rA}) into (\ref{G22rA}) and obtain 
\begin{equation}
D'' = \kappa P_z \textrm{e}^\mu D. \label{5t}
\end{equation}
Differentiating (\ref{5l}) with respect to $r$, we obtain $D''$ as a function of $\rho(r)$, $P_z(r)$, and their first and second derivatives, which we insert into (\ref{5t}) together with (\ref{5j}) and (\ref{5m}), and obtain
\begin{eqnarray}
&-& \kappa^2 P_z^3 (\rho^2 - P_z^2) \left(\sqrt{\frac{- \rho}{P_z}} - \sqrt{\frac{- P_z}{\rho}} \right) f^2 \pm c \sqrt{\frac{- P_z}{\rho}} \left(\frac{\rho'}{\rho} - \frac{P'_z}{P_z} \right) \rho^2 P_z \left(1 + \frac{P_z}{\rho}\right) \left\{ \left[ \frac{P''_z}{P_z} - \frac{\rho''}{\rho} + \left(\frac{\rho'}{\rho} + \frac{P'_z}{P_z} \right)\left(\frac{\rho'}{\rho} - \frac{P'_z}{P_z} \right) \right] \right. \nonumber \\
&& \left. \times \left(\sqrt{\frac{- \rho}{P_z}} + \sqrt{\frac{- P_z}{\rho}} \right) + \frac{1}{2} \left(\frac{\rho'}{\rho} - \frac{P'_z}{P_z} \right)^2  \left(\sqrt{\frac{- \rho}{P_z}} - \sqrt{\frac{- P_z}{\rho}} \right) \right\} f + c^2 P^2_z (\rho - P_z) \nonumber \\
&\times&\left\{ \left[ \frac{P''_z}{P_z} - \frac{\rho''}{\rho} + \left(\frac{\rho'}{\rho} + \frac{P'_z}{P_z} \right) \left(\frac{\rho'}{\rho} - \frac{P'_z}{P_z} \right) \right] \left(\sqrt{\frac{- \rho}{P_z}} + \sqrt{\frac{- P_z}{\rho}} \right) + \frac{1}{2}  \left(\frac{\rho'}{\rho} - \frac{P'_z}{P_z} \right)^2 \left(\sqrt{\frac{- \rho}{P_z}} - \sqrt{\frac{- P_z}{\rho}} \right) \right\} = 0. \label{5u}
\end{eqnarray}

Compared to (\ref{5q}), Eq. (\ref{5u}) implies, since each coefficient of, respectively, the $f^2$, the $f$ and the zero-order term must be proportional to each other in both equations,
\begin{eqnarray}
&-& \kappa^2 P^3_z  (\rho^2 - P_z^2) \left(\sqrt{\frac{- \rho}{P_z}} - \sqrt{\frac{- P_z}{\rho}} \right)\left[\frac{AC' - A'C}{A}\right]^{-1} = \pm  c \sqrt{\frac{- P_z}{\rho}} \left(\frac{\rho'}{\rho} - \frac{P'_z}{P_z} \right) \rho^2 P_z \left(1 + \frac{P_z}{\rho}\right) \nonumber \\
&& \times \left\{ \left[ \frac{P''_z}{P_z} - \frac{\rho''}{\rho} + \left(\frac{\rho'}{\rho} + \frac{P'_z}{P_z} \right)\left(\frac{\rho'}{\rho} - \frac{P'_z}{P_z} \right) \right] \left(\sqrt{\frac{- \rho}{P_z}} + \sqrt{\frac{- P_z}{\rho}} \right) + \frac{1}{2} \left(\frac{\rho'}{\rho} - \frac{P'_z}{P_z} \right)^2  \left(\sqrt{\frac{- \rho}{P_z}} - \sqrt{\frac{- P_z}{\rho}} \right) \right\} \nonumber \\
&\times& \left[\frac{AB' - A'B}{A} \pm 2c \left(\sqrt{\frac{- \rho}{P_z}} - \sqrt{\frac{- P_z}{\rho}} \right) C \right]^{-1} = \pm \left\{ \left[ \frac{P''_z}{P_z} - \frac{\rho''}{\rho} + \left(\frac{\rho'}{\rho} + \frac{P'_z}{P_z} \right) \left(\frac{\rho'}{\rho} - \frac{P'_z}{P_z} \right) \right] \left(\sqrt{\frac{- \rho}{P_z}} + \sqrt{\frac{- P_z}{\rho}} \right) \right. \nonumber \\
&& \left. + \frac{1}{2} \left(\frac{\rho'}{\rho} - \frac{P'_z}{P_z} \right)^2 \left(\sqrt{\frac{- \rho}{P_z}} - \sqrt{\frac{- P_z}{\rho}} \right) \right\} \left[2 \left(\sqrt{\frac{- \rho}{P_z}} - 2 \sqrt{\frac{- P_z}{\rho}} \right) \right]^{-1}, \label{5ua}
\end{eqnarray}
which is a double constraint on $\rho(r)$, $P_z$ and derivatives. These equations cannot simplify to mere trivial forms owing to the $\kappa^2$ factor in the first term which is present in neither of the two other ones. This $\kappa$ expression cannot even vanish, since it would imply either $P_z = 0$ or $P_z = \rho$, corresponding, repectively, to $\alpha = 0$ or $\alpha = 1$ in Appendix A where such eos are shown not to satisfy the sufficient conditions for purely magnetic spacetimes.

\subsection{Second solution of (\ref{5f})} \label{fs2}

This second solution reads
\begin{eqnarray}
2 \left(1 + \frac{P_z}{\rho} \right)ff'' + 2\left(\frac{\rho'}{\rho} - \frac{P'_z}{P_z} \right) ff' - \left(2 + \frac{P_z}{\rho} \right)f'^2+ 4c^2  \frac{P_z}{\rho} = 0. \label{5v}
\end{eqnarray}
Now, we differentiate (\ref{D2}) with respect to $r$ and obtain
\begin{equation}
l' = \frac{2(DD' - kk')}{f} - \frac{(D^2 - k^2) f'}{f^2}, \label{5w}
\end{equation}
which we insert into (\ref{G11rA}) together with (\ref{A1}) and  (\ref{5b}) and obtain
\begin{equation}
\frac{2 D'}{D} = \frac{f'^2 - 4 c^2}{\left(\frac{\rho}{P_z} +1\right) ff'}. \label{5x}
\end{equation}
The above results depend only on the field equations (and the Bianchi identity) for the considered (\ref{eosA}) eos. Now, we introduce the constraint that spacetimes should be purely magnetic in the sufficient sense displayed by (\ref{cond12b}) and (\ref{cond12e}).

Inserting (\ref{A1}) into (\ref{cond12b}), we obtain
\begin{equation}
D = c_5 f' \left(\frac{\rho}{P_z} - 1 \right), \label{5y}
\end{equation}
which we derive with respect to $r$, divide the result by (\ref{5y}) and obtain
\begin{equation}
\frac{2D'}{D} =  2 \frac{f''}{f'} + 2 \frac{P_z \rho' - \rho P'_z}{(\rho - P_z) P_z}, \label{5z}
\end{equation}
that we insert into (\ref{5x}), and the result into (\ref{5v}), that becomes
\begin{equation}
\frac{f'}{f} =  \frac{2 P_z}{P_z - \rho} \left(\frac{\rho'}{\rho} - \frac{P'_z}{P_z} \right), \label{5aa}
\end{equation}
which, used into (\ref{A1}), gives 
\begin{equation}
\mu' =  \frac{2 \rho}{P_z - \rho}\left(\frac{\rho'}{\rho} - \frac{P'_z}{P_z} \right). \label{5ab}
\end{equation}
Inserting (\ref{5x}), (\ref{5aa}) and (\ref{5ab}) into the (\ref{cond12e}) purely magnetic constraint equation where we have set $P_r=0$, we obtain
\begin{equation}
\textrm{e}^\mu = \frac{\kappa}{2} \frac{E f^2}{(G + H f^2)}, \label{5ac}
\end{equation}
with
\begin{equation}
E(r) = (\rho + P_z)^2 (\rho - P_z), \label{5ad}
\end{equation}
\begin{equation}
G(r) =  2c^2 \rho (\rho - P_z), \label{5ae}
\end{equation}
\begin{equation}
H(r) =  \rho P_z \left(\frac{\rho'}{\rho} - \frac{P'_z}{P_z} \right)^2. \label{5af}
\end{equation}
Notice that the regularity conditions are verified, identically by  (\ref{5y}), and by  (\ref{5ac}) provided
\begin{equation}
\frac{\kappa}{2} \frac{E}{(G + H)}  \stackrel{0}{=}  1. \label{5ae}
\end{equation}
Then, we differentiate (\ref{5ac}) with respect to $r$ and substitute (\ref{5aa}) and (\ref{5ab}) into the result, which gives
\begin{equation}
f^2 =  \frac{\left[ \frac{2(\rho - 2 P_z)}{\rho - P_z} \left(\frac{\rho'}{\rho} - \frac{P'_z}{P_z} \right) + \frac{E'}{E} \right] G - G'}{H' - \left[\frac{2 \rho}           {\rho - P_z} \left(\frac{\rho'}{\rho} - \frac{P'_z}{P_z} \right) + \frac{E'}{E} \right] H} . \label{5ag}
\end{equation}
Taking the square root of Eq. (\ref{5ag}), we choose the plus sign so that the metric signature should be consistent with our previous choice and obtain
\begin{equation}
f =  \left\{\frac{\left[ \frac{2(\rho - 2 P_z)}{\rho - P_z} \left(\frac{\rho'}{\rho} - \frac{P'_z}{P_z} \right) + \frac{E'}{E} \right] G - G'}{H' - \left[\frac{2 \rho}           {\rho - P_z} \left(\frac{\rho'}{\rho} - \frac{P'_z}{P_z} \right) + \frac{E'}{E} \right] H}\right\}^{1/2} . \label{5ah}
\end{equation}
The $f$ real requirement imposes the following constraint on $\rho(r)$, $P_z$, and their first and second derivatives:
\begin{equation}
\frac{\left[ \frac{2(\rho - 2 P_z)}{\rho - P_z} \left(\frac{\rho'}{\rho} - \frac{P'_z}{P_z} \right) + \frac{E'}{E} \right] G - G'}{H' - \left[\frac{2 \rho}{\rho - P_z} \left(\frac{\rho'}{\rho} - \frac{P'_z}{P_z} \right) + \frac{E'}{E} \right] H} > 0. \label{5ai} 
\end{equation}
Now, inserting (\ref{5ag}) into (\ref{5ac}), we obtain
\begin{equation}
\textrm{e}^\mu = \frac{\kappa E}{2 H} \left[1 +  \frac{\frac{H'}{H} -\frac{2 \rho}{\rho - P_z} \left(\frac{\rho'}{\rho} - \frac{P'_z}{P_z} \right) - \frac{E'}{E}}
{\frac{2(\rho - 2 P_z)}{\rho - P_z} \left(\frac{\rho'}{\rho} - \frac{P'_z}{P_z} \right) + \frac{E'}{E} - \frac{G'}{G}}\right]^{-1}. \label{5aj}
\end{equation}
Substituting $f'$, given by (\ref{5aa}) where we insert (\ref{5ah}), into (\ref{5y}), we obtain
\begin{equation}
D =  2 c_5 \left(\frac{P'_z}{P_z} - \frac{\rho'}{\rho} \right) \left\{\frac{\left[\frac{2(\rho - 2 P_z)}{\rho - P_z} \left(\frac{\rho'}{\rho} - \frac{P'_z}{P_z} \right) + \frac{E'}{E} \right] G - G'}{H' - \left[\frac{2 \rho}{\rho - P_z} \left(\frac{\rho'}{\rho} - \frac{P'_z}{P_z} \right) + \frac{E'}{E} \right] H}\right\}^{1/2} . \label{5ak}
\end{equation}
To obtain an equation for $k$, we insert (\ref{5y}) and (\ref{5aa}) into (\ref{5b}), which gives
\begin{equation}
k' +  \frac{2 P_z}{\rho - P_z} \left(\frac{\rho'}{\rho} - \frac{P'_z}{P_z} \right) k - \frac{4 c c_5}{f} \left(\frac{\rho'}{\rho} - \frac{P'_z}{P_z} \right) = 0, \label{5al}
\end{equation}
where we insert $f$ as given by (\ref{5ah}) and obtain a first order differential equation for $k$ that can be integrated as
\begin{equation}
k = \textrm{e}^{\int_{r_1}^r - W(u) \textrm{d}u} \left\{c_8  + \int_{r_1}^r Y(v) \left[ -  \textrm{e}^{\int_{r_1}^v W(u) \textrm{d}u} \right] \textrm{d}v \right\}, \label{5ala}
\end{equation}
where $c_8$ is an integration constant, $r_1$ is a constant integration limit verifying $r_1 < r_{\Sigma}$, and
\begin{equation}
W(r)  = \frac{2P_z}{\rho - P_z} \left(\frac{\rho'}{\rho} - \frac{P_z'}{P_z} \right), \label{5alb}
\end{equation}
\begin{equation}
Y(r)  = - \frac{4c c_5}{f}\left( \frac{\rho'}{\rho} - \frac{P_z'}{P_z} \right). \label{5alc}
\end{equation}
It is easy to see, considering (\ref{regcond4}) and (\ref{regcond7}), and since $c \neq 0$ and $c_5 \neq 0$, that  (\ref{5al}) is consistent with these regularity conditions provided
\begin{equation}
 \frac{\rho'}{\rho} - \frac{P_z'}{P_z} \stackrel{0}{=} 0, \label{5ald}
\end{equation}
which is a new regularity constraint on $\rho(r)$, $P_z(r)$ and their first derivatives, which has to be satisfied together with (\ref{5ae}). From both (\ref{5ae}) and (\ref{5ald}) regularity conditions, we obtain the simplified following one:
\begin{equation}
 \kappa (\rho + P_z)^2 \stackrel{0}{=} 4 c^2 \rho. \label{5ale}
\end{equation}
The last metric function $l(r)$ follows from (\ref{D2}), once $f(r)$, $k(r)$ and $D(r)$ are known.

As explained in Sec. \ref{fs1}, the differential equation number compared to the free $r$ function one implies a constraint equation on $\rho(r)$ and $P_z(r)$ which we derive again from (\ref{5t}). Differentiating twice (\ref{5y}) with respect to $r$ while inserting (\ref{5aa}) and substituting the result into (\ref{5t}) together with (\ref{5ac}) and (\ref{5ak}), we obtain
\begin{equation}
f^2 = \frac{- 2 G N}{2 H N + \kappa^2 P_z  \left(\frac{\rho'}{\rho} - \frac{P'_z}{P_z} \right) E}, \label{5am}
\end{equation}
with
\begin{eqnarray}
N &=& \frac{2}{\rho - P_z} \left(\frac{\rho'}{\rho} - \frac{P'_z}{P_z} \right) \left\{\frac{1}{\rho - P_z} \left(\frac{\rho'}{\rho} - \frac{P'_z}{P_z} \right) \left[- 2 P_z^2 \left(\frac{\rho'}{\rho} - \frac{P'_z}{P_z} \right) + \rho P'_z - P_z \rho' \right]  \right. \nonumber \\
&+& \left. 3 P_z \left(\frac{\rho''}{\rho} - \frac{\rho'^2}{\rho^2} - \frac{P''_z}{P_z} + \frac{P'^2_z}{P^2_z} \right) \right\} + \frac{P'''_z}{P_z} - \frac{3 P'_z P''_z}{P^2_z} + \frac{2 P'^3_z}{P^3_z} - \frac{\rho'''}{\rho} + \frac{3 \rho' \rho''}{\rho^2} -\frac{2 \rho'^3}{\rho^3}. \label{5an}
\end{eqnarray}
Now, we insert (\ref{5am}) into (\ref{5ag}) and obtain, after some algebra,
\begin{eqnarray}
&&\left\{\frac{4 P_z}{\rho - P_z}  \left(\frac{\rho'}{\rho} - \frac{P'_z}{P_z}\right) \left[- \frac{\rho + 2 P_z}{\rho - P_z} \left(\frac{\rho'}{\rho} - \frac{P'_z}{P_z} \right)^2 + 3 \left(\frac{\rho'}{\rho} - \frac{P'_z}{P_z} \right)' \right] - 2 \left(\frac{\rho'}{\rho} - \frac{P'_z}{P_z} \right)''\right\} \left[ \frac{\rho + 4 P_z}{\rho - P_z} \left(\frac{\rho'}{\rho} - \frac{P'_z}{P_z} \right)^2  \right. \nonumber \\
& -& \left. 2 \left(\frac{\rho'}{\rho} - \frac{P'_z}{P_z} \right)' \right] - \frac{\kappa^2 (\rho + P_z)^2(\rho - P_z)}{\rho} \left[2 \left(\frac{\rho'}{\rho} - \frac{P'_z}{P_z} \right) \frac{\rho^2 - 2 \rho P_z - P_z^2}{(\rho - P_z) (\rho + P_z)} + \frac{\rho'}{\rho} \right] =0, \label{5ao}
\end{eqnarray}
which is the constraint equation the fluid parameters must satisfy so that the corresponding spacetime, besides being a field equation solution, should also be purely magnetic. This constraint equation rapid examination shows it can be fulfilled by an infinite number of $\{\rho, P_z\}$ couples, one of them being the $P_z = \alpha \rho$ eos ruled out in Appendix A. However, nontrivial solutions do exist. As an existence proof, we display the following fully integrated solution class, leaving its property study and the search for other solutions to future work. An eos form leading to integrable physical parameters and metric functions is
\begin{equation}
\frac{P_z}{\rho} = h(r), \label{5ap}
\end{equation}
where $h(r)$ is a radial coordinate function satisfying the regularity condition (\ref{5ald}) by verifying $h'/h \stackrel{0}{=} 0$ . In this case, Eq. (\ref{5ao}) becomes the following first order ordinary differential equation for $\rho^2$:
\begin{eqnarray}
&&(\rho^2)' - 4\frac{(1 - 2h -h^2) h'}{(1-h^2)h} \rho^2 - \frac{2}{\kappa^2 (1+h)^2(1-h)} \left\{\frac{4h'}{1-h}\left[\frac{(1+2h)h'^2}{(1-h)h^2} + 3\left(\frac{h'}{h}\right)' \right] +2\left(\frac{h'}{h}\right)'' \right\} \nonumber \\
&\times&  \left[\frac{(1+4h)h'^2}{(1-h)h^2} + 2\left(\frac{h'}{h}\right)' \right] = 0, \label{5aq}
\end{eqnarray}
from which $\rho^2$ can be partially integrated as
\begin{eqnarray}
\rho^2 &=& \frac{h^4 (1-h)^4}{(1 + h)^4}  \left\{c_{9} + \frac{c_{10}}{\kappa^2} \int^r_{r_1} \frac{(1+h)^2}{(1-h)^5 h^4} \left\{\frac{4h'}{1-h} \left[ \frac{(1+2h) h'^2}{(1-h) h^2} +3\left(\frac{h'}{h} \right)' \right] +2\left(\frac{h'}{h}\right)'' \right\}\right. \nonumber \\
&\times& \left. \left[\frac{(1+4h)h'^2}{(1-h) h^2} + 2\left(\frac{h'}{h} \right)' \right] \textrm{d}v \right\},  \label{5ar}
\end{eqnarray}
where $c_{9}$ and $c_{10}$ are integration constants. This expression will be analytically or numerically fully integrated depending on the $h(r)$ function pertaining to the considered problem.

Now, exact metric function expressions for this eos class proceed from (\ref{D2}), (\ref{5y}), (\ref{5aa}), (\ref{5ab}), (\ref{5ala}), (\ref{5alb}) and (\ref{5alc}) where (\ref{5ap}) is inserted. They arise as fully integrated $h$ functions that read
\begin{equation}
\textrm{e}^{\mu}= c_{11}\frac{h^2}{(1-h)^2}, \label{5as}
\end{equation}
\begin{equation}
f = \frac{c_{12}}{(1-h)^2}, \label{5at}
\end{equation}
\begin{equation}
k= \frac{1}{(1-h)^2}\left[c_{13} -c_{14}\left( \ln h - 4h + 3h^2 - \frac{4}{3} h^3 +\frac{h^4}{4}\right) \right], \label{5au}
\end{equation}
\begin{eqnarray}
l = \frac{1}{(1 - h)^2}  \left\{c_{15} \frac{h'^2}{h^2} - \left[c_{16} - c_{17}\left(\ln h - 4h + 3h^2 - \frac{4}{3} h^3 + \frac{h^4}{4} \right) \right]^2 \right\}, \label{5av}
\end{eqnarray}
where $c_{11}$-$c_{17}$ are integration constants and which constitute an exact solution to Einstein's field equations for a purely magnetic interior spacetime exhibiting an eos form as given by (\ref{5ap}) and verifying, from the regularity and junction conditions, $h'/h \stackrel{0}{=} 0$ and $\rho(r=0) \neq 0$. This result does not preclude the existence of other eos classes verifying constraint equation (\ref{5ao}) and possibly exhibiting other exact solutions to the here analyzed issue.

\hfill

Hence, we have obtained, for the particular (\ref{eosA}) eos, two different solution classes for the five (\ref{G00rA}) - (\ref{G33rA}) field equations and the two (\ref{cond12b}) and (\ref{cond12e}) purely magnetic constraint equations. The first class implies $P_z < 0$ and the second allows $P_z>0$, which enables one to consider a standard fluid as the gravitational source. We have derived, for both solutions, $f(r)$, $k(r)$, $\textrm{e}^\mu(r)$, $l(r)$, and $D(r)$ as explicit $\rho(r)$, $P_z(r)$, and derivative expressions. We have also displayed in both cases one or two constraint equations that have to be satisfied by $\rho(r)$, $P_z(r)$, and derivatives so that the metric solutions are consistent with the whole equation set. Hence, given a couple $\{\rho(r), P_z(r)\}$ fulfilling such constraints, we have found exact solutions to the problem of deriving the metric for an interior purely magnetic spacetime sourced by a rigidly rotating stationary cylindrical anisotropic fluid exhibiting a (\ref{5ap}) eos.

Displaying a nontrivial fully integrated solution in (\ref{5ap}) - (\ref{5av}), we have explicitly shown that such purely magnetic interior spacetimes do exist. We are thus led to generalize this result to rigidly or nonrigidly rotating fluids exhibiting eos with more degrees of freedom and therefore able to fulfill more easily the constraints.

As an opposite example, we show in Appendix A that too simple an eos, such as the one proposed in \cite{D06}, even though a solution to the field equations can be characterized in this case, does not ensure obligatorily the corresponding spacetime to be Weyl purely magnetic, at least when using our $J=0$ and $I<0$ somehow stringent (\ref{vanC0303a}) and (\ref{vanC0102a}) conditions.

\section{Conclusions} \label{concl}

We have displayed a whole set of mathematical equations and physical elements characterizing the stationary cylindrically symmetric anisotropic fluid interior spacetimes for both nonrigidly and rigidly rotating fluids, therefore, in the last case, completing and improving results displayed in \cite{D06}. 

We have first established the metric, the field equations, the hydrodynamical scalars, vectors and tensors, the regularity and junction conditions, the stress-energy tensor conservation equation and conducted a gravito-electromagnetic analysis including the Weyl tensor and its three complex invariants. Using a rule demonstrated in \cite{M94} and taking exclusively these three invariants into account, we have thus been led to characterize first nonrigidly rotating fluid spacetime purely electric (magnetic) properties. In particular, we have proposed, as a strongly based conjecture, that purely electric and purely magnetic Petrov type $I(M^\infty)$ stationary cylindrical anisotropic nonrigidly rotating fluid interior spacetimes exist and we have displayed different simplified equation sets they can verify in each case, purely electric (three subcases) and purely magnetic (one subcase).

Then we have conducted an analogous analysis applied to rigid rotation which had already been considered in \cite{D06}. We have corrected a typo in their Eq. (12) and an inaccuracy appearing in their Sec. 5, and we have shown that, while purely electric cylindrically symmetric spacetimes are {\it not} necessarily static in the {\it nonrigid} rotation case, this remains an open question for rigid rotation.

Moreover, we have displayed an existence {\it proof} of purely magnetic Petrov type $I(M^{\infty})$ stationary cylindrical anisotropic fluid spacetimes of which we have established the determining equations. This result's importance comes from the fact that extremely few purely magnetic solutions are known, while most of the known ones are Petrov type $D$, less are Petrov type $I(M^\infty)$, and fewer are physically consistent. Hence, our new existence proof of purely magnetic stationary cylindrical anisotropic fluid interior spacetimes is of the utmost importance for the progress in understanding gravito-electromagnetism. We have shown that, in this cylindrically symmetric case, even shear-free fluid motion can source purely magnetic spacetimes, provided rotation should be involved. This statement obtains equally in the shearing nonrigid rotation case. 

Finally, we have found two new classes of exact solutions to Einstein's equation featuring purely magnetic interior spacetimes sourced by a rigidly rotating stationary cylindrical anisotropic fluid, characterized by a rather simple but physically consistent equation of state. Besides displaying new exact solutions to general relativity's field equations, this work has allowed us to characterize constraints to be fulfilled by the fluid parameter functions so that the new solutions actually  correspond to Weyl purely magnetic spacetimes. And, to provide an existence proof for solutions to such constraint equations, we have displayed a {\it fully integrated metric} corresponding to a large class of eos exhibiting the best physical motivations. Further study of these solution properties and of other special cases is left to future work.

Therefore, although purely magnetic {\it vacuum} spacetimes are widely thought not to exist, our results show that {\it nonvacuum} physically motivated such spacetimes do and might be of use for astrophysical purposes.

\appendix
\section{A counterexample}

As an example of spacetimes verifying the field equations but not our sufficient conditions for being purely magnetic, we now address the particular eos of Debbasch {\it et al.} \cite{D06} that we write in our notations as
\begin{equation}
P_z = \alpha \rho, \quad P_r = P_\phi = 0, \label{eos}
\end{equation}
which, inserted into (\ref{Bianchi3}), gives Eq. (46) of Ref. \cite{D06}, recalled here as
\begin{equation}
\alpha \mu' = \frac{f'}{f}, \label{deb1}
\end{equation}
which can be integrated as \cite{D06}
\begin{equation}
\textrm{e}^{\alpha \mu} = f. \label{deb2}
\end{equation}
With (\ref{eos}) inserted into the field equations (\ref{G00rA})-(\ref{G33rA}), while using  (\ref{D2}), (\ref{regcond4}), (\ref{partint7}), (\ref{c}), and (\ref{deb1}), we obtain, after integration and provided $\alpha \neq 0$ \cite{D06},
\begin{equation}
f = \left(1 - \frac{2 +\alpha}{4 \alpha c^2} f'^2 \right)^{(1 + \alpha)/(2 + \alpha)}. \label{deb3}
\end{equation}
The above (\ref{deb3}) result is issued only from the field equations written for the chosen (\ref{eos}) eos. Now, we add the (\ref{cond12b}) and (\ref{cond12e}) sufficient conditions to try and see whether any purely magnetic property could be  exhibited by the corresponding spacetimes.
First, we insert (\ref{eos}) into (\ref{cond12e}) and obtain
\begin{equation}
\mu' \textrm{e}^\mu \left(\frac{f'}{f} - 4 \frac{D'}{D} \right) = 2 \kappa (1 + \alpha) \rho. \label{pm1}
\end{equation}
Then, using (\ref{cond12b}), (\ref{deb1}), (\ref{deb2}), (\ref{deb3}), its derivative with respect to $r$,  and (\ref{pm1}), we obtain
\begin{equation}
(3 + \alpha)(2 + \alpha) f^{(2 + \alpha)/(1 + \alpha)} - \frac{\kappa}{2 c^2} (2 + \alpha)(1 + \alpha) \rho f^{2-1/\alpha} + 1 = 0. \label{pm2}
\end{equation}

However, we can also achieve, with the same equation set, another very different equation supposed to determine $f$, but inconsistent with (\ref{pm2}). We proceed as follows. Inserting $l'$, obtained by differentiating $l$ extracted from (\ref{D2}), and using (\ref{partint7}) and (\ref{deb1}) into (\ref{G11r}), we obtain
\begin{equation}
\frac{2D'}{D} = \frac{f'^2 - 4 c^2}{\left(1 + \frac{1}{\alpha}\right) f f'}. \label{pm3}
\end{equation}
Now, $D$ can merely proceed from (\ref{deb1}) inserted into (\ref{cond12b}), as
\begin{equation}
D = c_5 f' \left(\frac{1}{\alpha} - 1\right), \label{pm4}
\end{equation}
into which we insert $f'$ as given by  (\ref{deb3}), which implies an expression for $D$ as a $f$ function. We differentiate this equation with respect to $r$ and obtain $D'$ which we use to compute
\begin{equation}
\frac{2D'}{D} = \frac{2 + \alpha}{1 + \alpha} \frac{f'}{(f - f^{-1/(1 + \alpha)})}. \label{pm5}
\end{equation}
Then, we equalize (\ref{pm3}) with (\ref{pm5}) and obtain
\begin{equation}
 f^{(2 + \alpha)/(1 + \alpha)} + f^{-(2 + \alpha)/(1 + \alpha)} - 2= 0. \label{pm6}
\end{equation}
It is easy to see that (\ref{pm2}) and (\ref{pm6}) are inconsistent. Actually, there exist two solutions to (\ref{pm6}): $\alpha = -2$ which inserted into (\ref{pm2}) gives the wrong equality $1 = 0$, ruling out this first solution, and $f = 1$, which implies $k=l=0$ and $\textrm{e}^{\mu} = 1$, and therefore a 3D-Minkowski spacetime, ruling out this second solution too. Another reasoning using $h(r)=\alpha$ into (\ref{5as})-(\ref{5av}) can also lead to a Minkowski spacetime. This implies that the two (\ref{cond12b}) and (\ref{cond12e}) constraint equations are incompatible with the field equations of interior solutions sourced by a rigidly rotating cylindrical fluid exhibiting such an eos which cannot therefore be characterized as purely magnetic with our method. The reason is that the degrees of freedom allowed by the (\ref{eos}) eos are too small  to allow a physically relevant solution to emerge out of such an overdetermined equation set. However, we cannot state such an eos is actually ruled out as implying a purely magnetic spacetime since the  (\ref{cond12b}) and (\ref{cond12e}) constraint equations are mere sufficient but not necessary conditions for imposing purely magnetic features.


\begin{thebibliography}{10}

\bibitem {G09} J. B. Griffiths and J. Podolsk\'y, {\it Exact Spacetimes in Einstein's General Relativity}, Cambridge Monographs on Mathematical Physics (Cambridge University Press, Cambridge, England, 2009).
\bibitem {S09} H. Stephani, D. Kramer, M. MacCallum, C. Honselaers and E. Herlt, {\it Exact Solutions to Einstein's Field Equations}, Cambridge Monographs on Mathematical Physics (Cambridge University Press, Cambridge, England, 2009).
\bibitem {LC19} T. Levi-Civita, $ds^2$ einsteiniani in campi newtoniani. IX: L'analogo del potenziale logaritmico, {\it Rend. Accad. Lincei} {\bf 28}, 101 (1919).
\bibitem {L24} C. Lanczos, \"Uber eine station\"are Kosmologie in Sinne der Einsteinschen Gravitationstheories, {\it Z. Phys.} {\bf 21}, 73 (1924).
\bibitem {L32} T. Lewis, Some special solutions of the equations of axially symmetric gravitational fields, {\it Proc. R. Soc. A} {\bf 136}, 176 (1932).
\bibitem {C19} M.-N. C\'el\'erier, R. Chan, M. F. A. da Silva, and N. O. Santos, Translation in cylindrically symmetric vacuum, {\it Gen. Relativ. Gravit.} {\bf 51}, 149 (2019).
\bibitem {vS37} W. J. van Stockum, The gravitational field of a distribution of particles rotating about an axis of symmetry, {\it Proc. R. Soc. Edinburg} A {\bf 57}, 135 (1938).
\bibitem {B20} K. A. Bronnikov, N. O. Santos, and A. Wang, Cylindrical Systems in General Relativity, {\it Classical Quantum Gravity} {\bf 37}, 113002 (2020).
\bibitem {D06} F. Debbasch, L. Herrera, P. R. C. T. Pereira, and N. O. Santos, Stationary cylindrical anisotropic fluid, {\it Gen. Relativ. Gravit.} {\bf 38}, 
1825 (2006).
\bibitem {M53} A. Matte, Sur de nouvelles solutions oscillatoires des \'equations de la gravitation, {\it Can. J. Math.} {\bf 5}, 1 (1953).
\bibitem {B62} L. Bel, Les \'etats de radiation et le probl\`eme de l'\'energie en relativit\'e g\'en\'erale, {\it Cahiers de Physique} {\bf 16}, 59 (1962); Radiation states and the Problem of Energy in General Relativity, {\it  Gen. Relativ. Gravit.} {\bf 32}, 2047 (2000).
\bibitem {C14} L. F. O. Costa and J. Nat\'ario, Gravito-electromagnetic analogies, {\it Gen. Relativ. Gravit.} {\bf 46}, 1792 (2014).
\bibitem {L02} C. Lozanovski, Uniqueness properties of purely magnetic LRS perfect fluid spacetimes, {\it Classical Quantum Gravity} {\bf 19}, 6377 (2002).
\bibitem {B04} A. Barnes, Purely magnetic spacetimes, {\it Proceedings of the XXVII Spanish Relativity meeting, Alicante, Spain, 2003} edited by J. A. Miralles, J. A. Font and J. A. Pons (University Alicante Press, Alicante, 2004).
\bibitem {B89} A. Barnes and R. R. Rowlingson, Irrotational perfect fluids with a purely electric Weyl tensor, {\it Classical Quantum Gravity} {\bf 6}, 949 (1989).
\bibitem {L95} W. M. Lesame, P. K. S. Dunsby, and G. F. R. Ellis, Integrability conditions for irrotational dust with a purely electric Weyl tensor: A tetrad analysis, {\it Phys. Rev.} D {\bf 52}, 3406 (1995).
\bibitem {BMS96} M. \'A. G. Bonilla, M. Mars, J. M. M. Senovilla, C. F. Sopuerta, and R. Vera, Comment on: 'Integrability conditions for irrotational dust with a purely electric Weyl tensor: A tetrad analysis', {\it Phys. Rev.} D {\bf 54}, 6565 (1996).
\bibitem {vE97} H. van Elst, C. Uggla, W. M. Lesame, G. F. R. Ellis, and R. Maartens, Integrability of irrotational silent cosmological models, {\it Classical Quantum Gravity} {\bf 14}, 1151 (1997).
\bibitem {M99} M. Mars, 3 + 1 description of silent universes: a uniqueness result for the Petrov type I vacuum case, {\it Classical Quantum Gravity} {\bf 16}, 3245 (1999).
\bibitem {F01} J. J. Ferrando and J. A. S\'aez, Comments on purely electric Weyl tensors, {\it Reference Frames and Gravitomagnetism, Proceedings of the XXIII Spanish Relativity Meeting, Valladolid, Spain, 2000} edited by J. F. Pascual-S\'anchez, L. Flor\'ia, A. San Miguel and F. Vicente (World Scientific, Singapore, 2001) p.305.
\bibitem {V04} N. Van den Bergh and L. Wylleman, Silent universes with a cosmological constant, {\it Classical Quantum Gravity} {\bf 21}, 2291 (2004).
\bibitem {V05} N. Van den Bergh and L. Wylleman, Silent universes with a positive cosmological constant, {\it Int. J. Mod. Phys.} A {\bf 20}, 2316 (2005).
\bibitem {W06a}  L. Wylleman and N. Van den Bergh, Petrov type I silent universes with G3 isometry group: The uniqueness result recovered, {\it Classical Quantum Gravity} {\bf 23}, 329 (2006).
\bibitem {A94} R. Arianrhod, A. W.-C. Lun,  C. B. G. McIntosh and Z. Perj\'es, Magnetic curvatures, {\it Classical Quantum Gravity} {\bf 11}, 2331 (1994).
\bibitem {B95} W. B. Bonnor, The magnetic Weyl tensor and the van Stockum solution, {\it Classical Quantum Gravity} {\bf 12}, 1493 (1995).
\bibitem {M98} R. Maartens and B. A. Bassett, Gravito-electromagnetism, {\it Classical Quantum Gravity} 15, 705 (1998).
\bibitem {F99} G. Fodor, M. Marklund and Z. Perj\'es, Axistationary perfect fluid - a tetrad approach, {\it Classical Quantum Gravity} 16, 453 (1999).
\bibitem {LA99} C. Lozanovski and M. Aarons, Irrotational perfect fluid spacetimes with a purely magnetic Weyl tensor, {\it Classical Quantum Gravity} {\bf 16}, 4075 (1999).
\bibitem {LM99} C. Lozanovski and  C. B. G. McIntosh, Perfect fluid spacetimes with a purely magnetic Weyl tensor, {\it Gen. Relativ. Gravit.} {\bf 31}, 1355 (1999).
\bibitem {L03} C. Lozanovski and J. Carminati, Purely magnetic locally rotationally symmetric spacetimes, {\it Classical Quantum Gravity} {\bf 20}, 215 (2003).
\bibitem {W06b} L. Wylleman and N. Van den Bergh, Complete classification of purely magnetic, nonrotating, nonaccelerating perfect fluids, {\it Phys. Rev.} D {\bf 74}, 084001 (2006).
\bibitem {L07} C. Lozanovski, Szekeres-type mappings of Kasner and Petrov type $I(M^+)$ purely magnetic spacetimes, {\it Classical Quantum Gravity} {\bf 24}, 1169 (2007).
\bibitem {M94} C. B. G. McIntosh, R. Arianrhod, S. T. Wade, and C. Hoenselaers, Electric and magnetic Weyl tensors: classification and analysis, {\it Classical Quantum Gravity} {\bf 11}, 1555 (1994).
\bibitem {T65} M. Tr\"umper, On a special class of type-I gravitational fields, {\it J. Math. Phys.} (N.Y.) {\bf 6}, 584 (1965).
\bibitem {B75} C. H. Brans, Some restrictions on algebraically general vacuum metrics, {\it J. Math. Phys.} (N.Y.) {\bf 16}, 1008 (1975).
\bibitem {H95} B. M. Haddow, Purely magnetic space-times, {\it J. Math. Phys.} (N.Y.) {\bf 36}, 5848 (1995).
\bibitem {V03a} N. Van den Bergh, Purely gravito-magnetic vacuum spacetimes, {\it Classical Quantum Gravity} {\bf 20}, L1 (2003).
\bibitem {V03b} N. Van den Bergh, Tidal effects cannot be absent in a vacuum, {\it Classical Quantum Gravity} {\bf 20}, L165 (2003).
\bibitem {F03} J. J. Ferrando and J. A. S\'aez, Gravito-magnetic vacuum spacetimes: Kinematic restrictions, {\it Classical Quantum Gravity} {\bf 20}, 2835 (2003).
\bibitem {F04a} J. J. Ferrando and J. A. S\'aez, Aligned electric and magnetic Weyl fields, {\it Gen. Relativ. Gravit.} {\bf 36}, 2497 (2004).
\bibitem {F04b} J. J. Ferrando and J. A. S\'aez, On the classification of type D space-times, {\it J. Math. Phys.} (N.Y.) {\bf 45}, 652 (2004).
\bibitem {Z05} E. Zakhary and J. Carminati, On purely gravito-magnetic vacuum space-times, {\it Gen. Relativ. Gravit.} {\bf 37}, 605 (2005).
\bibitem {S68} J. M. Stewart and G. F. R. Ellis, Solutions of Einstein's equations for a fluid which exhibits local rotational symmetry, {\it J. Math. Phys.} (N.Y.) {\bf 9}, 1072 (1968).
\bibitem {vE96} H. van Elst and G. F. R. Ellis, The covariant approach to LRS perfect fluid spacetime geometries, {\it Classical Quantum Gravity} {\bf 13}, 1099 (1996).
\bibitem {B96} M. Bradley and M. Marklund, Finding solutions to Einstein's equations in terms of invariant objects, {\it Classical Quantum Gravity} {\bf 13}, 3021  (1996).
\bibitem {Ma97} M. Marklund, Invariant construction of solutions to Einstein's field equations -- LRS perfect fluids I , {\it Classical Quantum Gravity} {\bf 14}, 1267 (1997).
\bibitem {Ma99} M. Marklund and M. Bradley, Invariant construction of solutions to Einstein's field equations -- LRS perfect fluids II, {\it Classical Quantum Gravity} {\bf 16}, 1577 (1999).
\bibitem {M90} C. B. G. McIntosh and  R. Arianrhod, 'Degenerate' nondegenerate spacetime metrics, {\it Classical Quantum Gravity} {\bf 7}, L213 (1990).
\bibitem {A92}  R. Arianrhod and C. B. G. McIntosh, Principal null directions of Petrov type I Weyl spinors: Geometry and symmetry, {\it Classical Quantum Gravity} {\bf 9}, 1969 (1992).
\bibitem {H73} S. W. Hawking and G. F. R. Ellis, {\it The Large Scale Structure of Spacetime,} Cambridge Monographs on Mathematical Physics (Cambridge University Press, Cambridge, England, 1973).
\bibitem {dS95} M. F. A. da Silva, L. Herrera, F. M. Paiva and N.O. Santos, The parameters of the Lewis metric for the Weyl class, {\it Gen. Relativ. Gravit.} {\bf 27}, 859 (1995).
\bibitem {D27} G. Darmois, Les \'equations de la gravitation einsteinienne, {\it M\'emorial des Sciences Mathématiques} fasc. XXV (Gauthier-Villars, Paris, 1927).
\bibitem {P86} R. Penrose and W. Rindler, {\it Spinors  and Spacetime} Vol. II (Cambridge University Press, Cambridge, England, 1986).

\end{thebibliography}
\end{document}